\documentclass[apj, numberedappendix]{emulateapj}

\usepackage{graphicx} 
\usepackage{amsmath}
\usepackage{hyperref}
\usepackage{amsfonts}
\usepackage{amsmath}
\usepackage{amssymb}
\usepackage{amsthm}
\usepackage{subeqnarray}
\usepackage{epstopdf}

\newcommand{\p}{\partial}

\newcommand{\brak}[1]{\langle #1\rangle}

\newcommand{\AU}{\; {\rm AU}}
\newcommand{\yr}{\; {\rm yr}}

\DeclareMathSymbol{\varOmega}{\mathord}{letters}{"0A}
\DeclareMathSymbol{\varSigma}{\mathord}{letters}{"06}
\DeclareMathSymbol{\varPsi}{\mathord}{letters}{"09}

\newcommand{\Eq}[1]{Equation\,(\ref{#1})}
\newcommand{\Eqs}[2]{Equations (\ref{#1}) and~(\ref{#2})}

\newcommand{\Eqsss}[3]{Equations (\ref{#1}), (\ref{#2}) and~(\ref{#3})}
\newcommand{\App}[1]{Appendix~\ref{#1}}

\newcommand{\Fig}[1]{Figure~\ref{#1}}
\newcommand{\Figs}[2]{Figs.~\ref{#1} and \ref{#2}}

\newcommand{\delad}{\nabla_{\rm ad}}
\newcommand{\delrad}{\nabla_{\rm rad}}
\newcommand{\Rg}{\mathcal{R}}
\newcommand{\RB}{R_{\rm B}}
\newcommand{\RH}{R_{\rm H}}
\newcommand{\co}{_{\rm c}}
\newcommand{\pla}{_{\rm p}}
\newcommand{\di}{_{\rm d}}
\newcommand{\cb}{_{\rm RCB}}
\newcommand{\mc}{m_{\rm c \oplus}}
\newcommand{\mcn}[1] { m_{ \rm c #1} }
\newcommand{\mpn}[1] { m_{ \rm p #1} }
\newcommand{\MC}{M_{\rm crit}}
\newcommand{\au}{a_\oplus}
\newcommand{\aun}[1]{ a_{#1} }

\begin{document}

\shortauthors{Piso \& Youdin}

\title{On the Minimum Core Mass for Giant Planet Formation at Wide Separations}

\author{Ana-Maria A. Piso}
\affil{Harvard-Smithsonian Center for Astrophysics, 60 Garden Street, Cambridge, MA 02138}

\author{Andrew N. Youdin}
\affil{JILA, University of Colorado and NIST, 440 UCB, Boulder, CO 80309-0440}

\begin{abstract}

In the core accretion hypothesis, giant planets form by gas accretion onto solid protoplanetary cores.  The minimum (or critical) core mass to form a gas giant is typically quoted as $10 M_{\oplus}$. The actual value depends on several factors: the location in the protoplanetary disk, atmospheric opacity, and the accretion rate of solids. Motivated by ongoing direct imaging searches for giant planets, this study investigates core mass requirements in the outer disk.  To determine the fastest allowed rates of gas accretion, we consider solid cores that no longer accrete planetesimals, as this would heat the gaseous envelope. Our spherical, two-layer atmospheric cooling model includes an inner convective region and an outer radiative zone that matches onto the disk.  We determine the minimum core mass for a giant planet to form within a typical disk lifetime of 3 Myr.   The minimum core mass declines with disk radius, from $\sim$$8.5 M_{\oplus}$ at 5 AU to $\sim$$3.5 M_{\oplus}$ at 100 AU, with standard interstellar grain opacities.  Lower temperatures in the outer disk explain this trend, while variations in disk density are less influential.  At all distances, a lower dust opacity or higher mean molecular weight reduces the critical core mass. Our non-self-gravitating, analytic cooling model reveals that self-gravity significantly affects early atmospheric evolution, starting when the atmosphere is only $\sim$$10\%$ as massive as the core.

\end{abstract}

\section{Introduction}
\label{intro}

Models of giant planet formation fall in two categories:  core accretion or gravitational instability \citep{dangelo11, youdin13}. In core accretion models, a solid core grows until it becomes massive enough to rapidly accrete gas \citep{PerCam74, mizuno78}.  Gravitational instability (GI) theories investigate the fragmentation of the protoplanetary disk into bound clumps \citep{cameron78, boss97}. 

Forming giant planets at wide separations in the protoplanetary disk poses theoretical challenges for both models.  While disks cannot fragment close to their host star, GI is more promising in the outer disk \citep{matzner05, rafikov05}.   However,  for GI to form planetary mass objects versus brown dwarfs, both instantaneous disk conditions and the history of gas infall must be finely tuned \citep{kratter10}.  The main concern for core accretion models, which are successful in the inner disk, is that they operate too slowly in the outer disk. The few Myr lifetime of gas disks sets the constraint \citep{williams11}.  However, mechanisms for the rapid accretion of solids, even at large separations, exist \citep{dones93, kenyon09, lambrechts12}. We thus consider the complementary problem of gas accretion timescales in the outer disk.

Core accretion models fall in several categories: static, quasi-static evolutionary and hydrodynamic.
In static models, the accretion of planetesimals provides a steady luminosity that determines the structure and mass of the atmosphere \citep{stevenson82}.    For a given planetesimal accretion rate, static solutions only exist up to a maximum core mass, referred to as the ``critical core mass",  $\MC$. Above this mass, the atmosphere is assumed to collapse and rapidly accrete disk gas, a process known as the ``core accretion instability."

Many static studies reproduce a canonical $\MC \sim 10 M_\oplus$.  However, \citet[hereafter R06]{Raf11, rafikov06} finds a wide range of values, $0.1 M_\oplus \lesssim \MC \lesssim 100 M_\oplus$, depending on the planetesimal accretion rate and disk parameters. 
Section \ref{sec:placc} compares our results to R06.  A general limitation of static models is the neglect of heat generated by atmospheric collapse, which can transform the core accretion instability into a slower process of Kelvin-Helmholtz (KH) contraction.

Quasi-static models include time-dependent atmospheric evolution due to KH contraction, and typically include a time-varying planetesimal accretion rate \citep{boden86, alibert05}.   Successful quasi-static models describe three phases of giant planet formation \citep{pollack96}.  In phase 1, rapid planetesimal accretion gives significant core growth, but prevents the accretion of a massive atmosphere.  When planetesimal accretion abates, as the core's feeding zone is depleted of solids, phase 2 begins.  During phase 2, the atmosphere grows gradually, cooling by KH contraction.  An ongoing trickle of planetesimal accretion can heat the atmosphere and slow this contraction.  Eventually, the atmosphere reaches the ``crossover mass", when it equals the mass of the now slowly growing core.  Phase 3, the runaway growth of the atmosphere, begins around the crossover mass.  This runaway occurs quasi-statically, i.e.\ in hydrostatic balance, but very rapidly compared to disk lifetimes. 

Dynamical models are needed to understand how runaway growth ends and final planet masses are determined.  Relevant processes -- including gap opening in disks and the transition of accretion from spherical to planar -- can be simulated and then added to atmospheric evolution calculations \citep{LisHub09}.

This work develops quasi-static models in which the core has a fixed mass and is no longer accreting solids.  By isolating the process of KH contraction, we determine the minimum timescale for runaway atmospheric growth, and thus for giant planet formation by core accretion.  Previous studies have considered this type of growth \citep[hereafter I00; PN05, respectively]{ikoma00, pn05}, which is an end-member case of phases 2 and 3 in more detailed quasi-static models.  Especially since phase 2 of atmospheric growth is often the slowest stage of core accretion, understanding the fastest allowed rates of gas accretion is of crucial importance.  Our study develops simplified cooling models for the purposes of broadly exploring parameter space -- especially in the outer disk -- and developing a more detailed understanding of atmospheric accretion.

This paper is organized as follows. Section \S\ref{sec:model} describes our model of atmospheric growth.  In Section \S\ref{sec:coolingan}, we develop a simplified analytic version of our atmospheric model.  We describe the structure and evolution of our atmosphere solutions in Section \S\ref{sec:KH}.   Section \S\ref{sec:critical} gives our final results for growth timescales and critical core masses.  We discuss some neglected effects in Section \S\ref{sec:neglected} and summarize our findings in Section \S\ref{sec:conclusions}.  Some detailed derivations are presented in the appendices.

\section{Atmospheric Accretion Model} \label{sec:model}

To model the accretion of gas by a solid protoplanet, we develop a simplified two-layer model for time-dependent atmospheric growth via cooling, i.e. KH contraction.  With a convective interior and radiative exterior, this model is motivated by similar models of hot Jupiters \citep{ab06, ym10}. 

Our model can  treat atmospheric growth up to the early stages of runaway growth, around the crossover mass.  At higher masses, our approximate treatment of the radiative zone (explained below) breaks down.   Since evolution after the onset of runaway growth contributes minimally to the total planet formation timescale, we can model growth times to good accuracy. 

Our main assumptions are summarized as follows:
\begin{enumerate}
\item The atmosphere is spherically symmetric, remains in hydrostatic balance, and matches onto the disk's midplane temperature and pressure at the planet's Hill radius.
\item The core mass and radius are fixed in evolutionary calculations, neglecting ongoing planetesimal or dust accretion.
\item Gravitational contraction of the atmosphere is the only source of planetary luminosity.  
\item Cooling of the atmosphere is dominated by the convective interior.  Thus the luminosity in the radiative exterior is held spatially constant. We justify this assumption in \S\ref{sec:twolayer} and confirm its validity in \S\ref{sec:endoftime}.
\item The atmosphere obeys a polytropic equation of state (EOS), with adiabatic index $\gamma = 7/5$ for an ideal diatomic gas.  Corrections from a realistic EOS are discussed in \S\ref{sec:EOS} and deferred to future work.
\item Dust grains provide the opacity in the radiative zone. For the cool temperatures in the outer disk, the radiative layer remains cool enough to avoid dust sublimation. Other opacity choices are discussed in \S\ref{sec:op}.
\end{enumerate}

The first assumption of spherical accretion breaks down early in the inner disk, as explained below.  However, our focus is on the outer disk, where this standard approximation is better justified.

\subsection{Disk and Opacity Model}\label{sec:disk}

We adopt a minimum mass solar nebula (MMSN) model for a passively irradiated disk \citep{chiang10}. With the semimajor axis $a$ normalized to the outer disk as $\aun{10} = a/(10 \text{ AU})$, the gas surface density and mid-plane temperature are  
\begin{subeqnarray} \label{eq:diskparam}
\varSigma\di  &=& 70 \,F_\varSigma \aun{10}^{-3/2} ~{\rm g~cm}^{-2} \\
T\di &=& 45  \,F_T\, \aun{10}^{-3/7} ~{\rm K} \, .
\end{subeqnarray}
The normalization factors $F_{\Sigma}$ and $F_T$  adjust the model relative to the fiducial MMSN.  We fix $F_{\varSigma}=F_T=1$ unless noted otherwise.  Some observations support a flatter surface density profile, $\varSigma\di \propto a^{-1}$ \citep{andrews10}.  We find that gas density and pressure are weak corrections to atmospheric growth (see Section \S\ref{sec:critical}).  However, a greater surface density of solids in the outer disk could favor the rapid growth of cores \citep{bromley11}.

For a vertically isothermal disk in hydrostatic balance (with no self-gravity), the mid-plane pressure of disk gas is 
\begin{equation}
\label{eq:Pd}
P\di = 6.9 \times 10^{-3} F_\varSigma \sqrt{F_T} \, \aun{10}^{-45/14}~{\rm dyn~cm^{-2}}
\end{equation}
for a molecular weight of $\mu=2.35$ proton masses and a Solar mass star.  

The (thermodynamically isothermal) sound speed in the disk is
\begin{equation}
c\di = \sqrt{\Rg T\di} = 0.4 \sqrt{F_T} \aun{10}^{3/14} ~\text{km s}^{-1} \,,
\end{equation}  
in terms of the specific gas constant $\Rg$.  The disk scale height is 
\begin{equation}
H\di = {c\di / \varOmega} = 0.42 \sqrt{F_T}  \, \aun{10}^{9/7} \AU\, ,
\end{equation} 
in terms of the Keplerian frequency $\varOmega = \sqrt{G M_\ast/a^3}$, with $G$ the gravitational constant and $M_\ast$ the stellar (in this work Solar) mass. 

We assume a dust opacity characteristic of interstellar grains, following \citet{bell94}:
\begin{equation}
\label{eq:opacitylaw}
\kappa= 2 F_\kappa  \left(\frac{T}{100\; \rm{K}}\right)^{\beta} \; \mathrm{cm^2 ~ g^{-1}},
\end{equation}
with a power-law index $\beta = 2$ and normalization $F_\kappa = 1$ unless noted otherwise. Grain growth tends to lower both $F_\kappa$ and $\beta$, while dust abundance scales with $F_\kappa$.   Section \S\ref{sec:op} discusses dust sublimation and more realistic opacity laws.

\subsection{Length Scales}
\label{sec:scales}

The characteristic length scales for protoplanetary atmospheres are crucial for choosing boundary conditions and for understanding the validity of  spherical symmetry in a gas disk of scale height $H\di$.

The solid core has a radius
\begin{equation}
\label{eq:rc}
R\co \equiv \left(\frac{3 M\co}{4 \pi \rho\co}\right)^{1/3} \approx 10^{-4} \mcn{10}^{1/3} ~\text{AU},
\end{equation}
where the core mass, $M\co$, is normalized to 10 Earth masses as $\mcn{10} \equiv M\co/(10~M_\oplus)$. The core density is held fixed at $\rho\co=3.2$ g cm$^{-3}$, representing a mixture of ice and rocky material \citep{pap99}.  We thus neglect  the detailed EOS of the solid core \citep{fortney07}.

A planet can bind a dense atmosphere if its escape velocity exceeds the sound speed.  This criterion is satisfied inside the Bondi radius
\begin{equation}
\label{eq:RB}
\RB \equiv \frac{G M\pla}{c\di^2} \approx 0.17 \, {\mpn{10}  \, \aun{10}^{3/7} \over F_T} ~\AU,
\end{equation}
where the enclosed planet mass, $M\pla = M\co + M_\mathrm{atm}$, includes the core and any atmosphere within the Bondi radius and $\mpn{10} \equiv M\pla /(10~M_\oplus)$.   The contribution of the atmospheric mass is small in early evolutionary stages.

Stellar tides dominate the planet's gravity beyond the Hill radius
\begin{equation}
\label{eq:RHill}
R_{\rm H} = \left(M\pla \over 3 M_\ast \right)^{1/3}a \approx 0.22 \, {\mpn{10}^{1/3} \, \aun{10} }~\AU ,
\end{equation}
where spherical symmetry and hydrostatic balance break down.  (Here we use the same symbol, $M\pla$, to now mean the mass within $\RH$.) 

The relevant length scales of the atmosphere and disk satisfy the relation $\RB H\di^2 = 3 R_{\rm H}^3$.  The length scales are roughly equal at the ``thermal mass'' (e.g., \citealt{menou04})
\begin{equation}\label{eq:Mth}
M_{\rm th} > {c\di^{3} \over G \varOmega} \approx 25 \, {F_T^{3/2} \over \sqrt{m_\ast} } \, \aun{10}^{6/7}~ M_\oplus \, .
\end{equation} 

In the low mass regime, $M\pla < M_{\rm th}/\sqrt{3}$, the length scales order as $\RB< \RH<H\di$.  In this regime, many studies assume the atmosphere matches the disk conditions at $\RB$.  We use $\RH$ as the matching radius, i.e.\ outer boundary, in all regimes. We discuss this choice further below.  


For a finite range of intermediate masses, $M_{\rm th}/\sqrt{3} < M\pla < 3 M_{\rm th}$, the Hill radius is the smallest scale, satisfying both $\RH < \RB$ and $\RH < H\di$.  Spherical symmetry remains a good, if imperfect, approximation because the disk is only weakly vertically stratified on  scales $\lesssim H\di$.  

At higher planet masses where $M\pla > 3 M_{\rm th}$ and $H\di < \RH < \RB$, spherical symmetry is no longer a good approximation, due to both the vertical stratification of the disk and gap opening.  See \S\ref{sec:hydro} for further discussion of neglected hydrodynamic effects.

We quote planet masses as the enclosed mass within the smaller of $\RB$ and $\RH$.  Thus when $\RB < \RH$, our computational domain -- which always ends at $\RH$ -- extends further than the location where we define atmosphere mass, within $\RB$.  We emphasize that this mass definition does not affect the evolutionary calculation, which consistently includes the enclosed mass at all radii.   We justify our mass convention as follows.  First, we want to conservatively define planet mass in a way that does not exaggerate atmospheric growth and is more consistent with previous works that use $\RB$ as the outer boundary in this regime (e.g., \citealt{ikoma00}, \citealt{pollack96}).  Second, most of the gas outside $\RB$ is weakly bound, and unlikely to remain attached to the planet if the disk suddenly dissipates.

The choice of $\RH$ as the outer boundary is based on this being the largest radius where spherical hydrostatic balance is approximately (but not exactly, see \S\ref{sec:hydro}) valid.  When $\RB < \RH$, we thus include the fact that the density at $\RB$ slightly exceeds the background disk density (see R06).  In practice, this effect is not very significant.  We would get similar results by choosing the outer boundary at $\RB$ in this regime, as previous studies have done.


%

\subsection{Structure Equations and Boundary Conditions}
\label{sec:struct}

Our atmosphere calculations use the standard structure equations of mass conservation, hydrostatic balance, thermal gradients, and energy conservation:
\begin{subeqnarray}
\label{eq:struct}
\frac{dm}{dr}&=&4 \pi r^2 \rho\slabel{eq:structb} \\
\frac{dP}{dr}&=&-\frac{G m}{r^2}\rho \slabel{eq:structa} \\
\frac{dT}{dr}&=&\nabla \frac{T}{P}\frac{dP}{dr}\slabel{eq:structc} \\
\frac{dL}{dr}&=&4 \pi r^2 \rho \left(\epsilon - \left. T {\partial S \over \partial t} \right|_m \right)\slabel{eq:structd}, 
\end{subeqnarray}
\noindent where $r$ is the radial coordinate, $L$ is the luminosity, and $P$, $T$, $\rho$  and $S$ are the gas pressure, temperature, density and entropy, respectively.  The enclosed mass  at radius $r$ is $m$. \Eq{eq:structc} simply defines the temperature gradient  $\nabla \equiv d \ln T/d \ln P$.  

Radiation diffusion gives a temperature gradient
\begin{equation}
\label{eq:delrad}
\delrad \equiv \frac{3 \kappa P}{64 \pi G m \sigma T^4} L\, ,
\end{equation}
with $\sigma$ the Stefan-Boltzmann constant.  In our models, optically thick diffusion is a good approximation throughout the radiative zones.  In convectively unstable regions, efficient convection gives an isentropic temperature gradient with $\nabla = \delad$, the adiabatic gradient 

\begin{equation}
\label{eq:delad}
\delad \equiv \Big(\frac{d \ln T}{d \ln P}\Big)_{\rm{ad}}.
\end{equation}
According to the Schwarzschild criterion, convective instability occurs when $\delrad > \delad$.  Thus $\nabla = \min(\delrad, \delad)$ sets the temperature gradient.

In the energy equation (\ref{eq:structd}), $\epsilon$ represents all local sources of heat input, except for the motion of the atmosphere itself.  From stellar structure, $\epsilon$ may be familiar as a nuclear burning term.  In a protoplanetary atmosphere, dissipative drag on planetesimals contributes to $\epsilon$.  Our models set $\epsilon = 0$, consistent with our neglect of planetesimal accretion.

The energy input from gravitational contraction, $\epsilon_{\rm g} = -T \partial S / \partial t$, is crucial for a cooling model.\footnote{In general, any motion is accounted for by this term.  The partial time derivative is performed on shells of fixed mass.}  The partial time derivative would normally require our radial derivative to be partial as well.  However, our subsequent developments will replace the local energy equation (\ref{eq:structd}) with global energy balance (see section \S\ref{cooling}), reverting the structure equations to time-independent ordinary differential equations (ODEs). 

To solve the equation set (\ref{eq:struct}), an EOS is required for closure. In our study, we adopt an ideal gas law with a polytropic EOS 
\begin{subeqnarray}\label{eq:idealEOS}
P &=& \rho \Rg T \, ,\slabel{eq:idealgas} \\
S &=& \Rg \ln \left(T^{1/\delad} \over P \right) \, ,\slabel{eq:polyEOS}  
\end{subeqnarray}
with $S$ a relative entropy.  This work uses $\delad=2/7$, i.e.\  a polytropic index $\gamma \equiv 1/(1 - \delad) = 7/5$, for an ideal diatomic gas.\footnote{\Eq{eq:polyEOS} is equivalent to the standard polytropic relation,  $P =K \rho^{\gamma}$, if the polytropic index $K$ replaces $S$.}    We thus neglect the presence of monatomic Helium in our choice of polytropes, a common practice in idealized studies.  We do, however, account for Helium in our reference value of $\mu = 2.35$ proton masses. 

Boundary conditions must be satisfied at both the base and the top of the atmosphere, with $m(R\co) = M\co$, $T(\RH) = T\di$ and $P(\RH) = P\di$.  Our model atmospheres also obey $L(R\co) = 0$. Along with \Eq{eq:structd}, this boundary condition is incorporated in the global cooling model described below.

\subsection{Global Cooling of an Embedded Planet}\label{cooling}

We now consider the global energy balance of a planet embedded in a gas disk.  More generally, our derivation applies to any spherical, hydrostatic object in pressure equilibrium with a background medium.  The total atmospheric energy includes gravitational and internal energies, $E = E_G + U$, with
\begin{subeqnarray}
E_G&=&-\int_{M\co}^M \frac{G m}{r} dm \, , \label{eq:Eg} \\
U&=&\int_{M\co}^M u dm \, .\slabel{eq:U}
\end{subeqnarray}
The specific internal energy is $u = C_V T = \Rg (\delad^{-1} -1) T$ for a polytropic EOS.  For a star or coreless planet, $M\co = 0$.

We start with a standard result, global energy balance for an isolated, i.e.\ not embedded, planet:
\begin{equation}
\label{eq:coolingstar}
L_M = L\co + \Gamma - \dot{E}.
\end{equation}
The surface luminosity, $L_M$, includes the core luminosity $L\co$ from e.g.\ planetesimal accretion or radioactive decay.   The total heat generation $\Gamma$  is the integral of $\epsilon$ over the atmosphere.  The rate of change of atmospheric energy, $\dot{E}$, is a loss term. 

For an object with no core luminosity (or no core) and no internal heat sources, the energy equation $L_M = -\dot{E}$ describes KH contraction in its simplest form.  When internal heat sources dominate, $L_M = \Gamma$, e.g.\ for nuclear burning in a main sequence star.

A protoplanetary atmosphere embedded in a gas disk lacks a free surface.  For objects without a free surface (or interior to a free surface), the full energy equation, 
\begin{equation}
\label{eq:coolingglobal}
L_M=L\co+\Gamma-\dot{E}+e_M \dot{M} - P_M \left. \frac{\partial V_M}{\partial t}\right|_M \, ,
\end{equation}
acquires surface terms as derived in  \App{sec:globalderiv}.  The surface can be at any mass level $M$, where the instantaneous radius is $R$.  Surface quantities are labeled by $M$ subscripts (except for $R$).  The energy accreted across the surface is given by the specific energy, $e_M = u_M-G M/R$, and the mass accretion rate of gas, $\dot{M}$.  The work done by the surface is $P_M \partial V_M/ \partial t$, with the partial derivative performed at fixed mass.  

For static solutions, which are not the focus of this paper, the surface terms (and also $\dot{E})$ vanish.  Static solutions are valid when imposed heat sources, i.e.\ $L\co$ and $\Gamma$, exceed the atmospheric losses.  Quantitatively, static solutions apply when the KH timescale,
\begin{equation}
\tau_{\rm KH} \sim {|E| \over L_M}, 
\end{equation} 
is shorter than the actual evolutionary timescale.  Thus, $\tau_{\rm KH}$, which our models calculate, gives strong lower limits on the time to form giant planets by core accretion.

\subsection{The Two-Layer Model} \label{sec:twolayer}

To simplify our evolutionary calculations, we develop a two-layer atmospheric model with a convective interior and a radiative exterior.   The existence of this layered structure is well known from previous studies (e.g., R06) and can be readily understood.  Before the protoplanetary atmosphere can cool, it has the entropy of the disk.  As the atmosphere cools, the deep interior remains convective.  Convective interiors are a common feature of low mass cool objects (brown dwarfs and planets) that results from the behavior of $\delrad$ for realistic opacity laws.  However, the entropy of the deep interior decreases as the atmosphere cools.  A region of outwardly increasing entropy, i.e.\ a radiative layer, is required to connect the convective interior to the disk.  A more complicated structure, with radiative windows in the convection zone, is possible as discussed in \S\ref{sec:op}. 

In convective regions, the adiabatic structure is independent of luminosity and can be calculated without local energy balance, \Eq{eq:structd}.  Thus, for fully convective objects, a cooling sequence can be established by connecting a series of adiabatic solutions using a global energy equation, $L_M = -\dot{E}$ or \Eq{eq:coolingstar}.  Such methods are commonly used for their computational efficiency and are sometime referred to as ``following the adiabats," since the steady state solutions evolve in order of decreasing entropy \citep{marleau13}.

In the radiative zone,  local energy balance, \Eq{eq:structd}, does affect the atmospheric structure.  We proceed by assuming that the majority of energy is lost from the convective interior, and thus the luminosity can be treated as constant in the outer radiative zone, i.e. the RHS of \Eq{eq:structd} is set to zero.  This assumption greatly simplifies the numerical problem.  Instead of solving time dependent partial differential equations, we can solve for  static solutions to a set of ODEs, Equations (\ref{eq:struct}a -- c), which we connect in a time series as described below. We show in \S\ref{sec:endoftime} that the approximation of constant luminosity in the outer layer holds for our regime of interest.   


To obtain a single atmosphere solution (indexed by $i$), we choose a planet mass $M_i$.  At the outer boundary, at $\RH(M_i)$, the temperature and pressure are set to the disk values.  A luminosity value is required to compute $\delrad$ and integrate Equations (\ref{eq:struct}a--c).  The correct value of the luminosity is not known in advance, and is the eigenvalue of the problem.  The boundary conditions can only be satisfied for the correct value of the luminosity eigenvalue, which we find by the shooting method.  Specifically, we alter the luminosity until the integrated value of mass at the core, $m(R\co)$, matches the actual core mass, $M\co$. Physically, the luminosity determines the location of the radiative-convective boundary (RCB), consistent with the structure equations and the Schwarzschild criterion.

To understand time evolution, we construct an array of solutions to Equations (\ref{eq:struct}a -- c) in order of increasing atmospheric mass.  We then use global energy balance, \Eq{eq:coolingglobal}, to ``follow the mass" and place these solutions in a cooling sequence.  To establish the time difference between neighboring solutions, we apply \Eq{eq:coolingglobal} at the RCB.  In principle, energy balance could be evaluated at any level.  Our approximate treatment of the radiative zone makes the RCB  the preferred location.     The elapsed time $\Delta t$  between states $i$ and $i +1$ is given by the finite difference
\begin{equation}
\label{eq:dti}
\Delta t = 
{ -\Delta E + \brak{e} \Delta M - \brak{P}  \Delta V_{\brak{M}} \over \brak{L} }\, ,
\end{equation} 
using \Eq{eq:coolingglobal} with $\Gamma = L\co = 0$.
Brackets indicate an average of, and $\Delta$ indicates a difference between, the two states.  All values are evaluated at the RCB.  Due to the partial derivative in \Eq{eq:coolingglobal}, the volume difference $\Delta V_{<M>}$ is performed at fixed mass, here the average of the masses at the RCB.

\section{Analytic Cooling Model}
\label{sec:coolingan}

This section develops the analytic version of our two-layer model. This analytic model is less accurate than our numerical model, primarily because it neglects self-gravity.  We show (in \S\ref{sec:KH}) that self-gravity becomes important at rather low atmospheric masses, $M_{\rm atm} \gtrsim 0.1 M\co$.  Nevertheless, the analytic model is useful for understanding atmospheric evolution and interpreting the numerical results.

The analytic model also assumes that the upper radiative layer is thick enough that $P\cb \gg P\di$.  This approximation ignores the earliest stages of cooling, which are rapid enough to be of minor importance.  The analytic model also ignores the surface terms in \Eq{eq:coolingglobal}, which we show to be a modest correction in \S\ref{sec:endoftime}.  Finally, while the numerical models set the outer boundary at $\RH$, we simplify the analytic calculations by setting the outer boundary to infinity, effectively neglecting the finite scale-height of the disk.   As shown in \App{iso}, the effect of this approximation is minor for $\RB \lesssim \RH$, the primary case of interest (see also R06).

Of all the approximations, the neglect of self-gravity is by far the most significant, as we have verified by comparison to numerical integrations that only neglect self-gravity.

\subsection{Two Layer Structure}
In order to apply the two-layer cooling model analytically, we require expressions for the atmospheric structure.  Conditions at the RCB are crucial as they set the interior adiabat and the radiative losses from the interior. (Recall that luminosity generation in the radiative zone is neglected.)  We express the temperature and pressure of the RCB, at the radius $R\cb$, as 
\begin{subeqnarray}\label{eq:cb2}
T\cb &=& \chi T\di \slabel{eq:Tcb} \\
P\cb &=& \theta P_{\rm d} e^{R_{\rm B}/R\cb} \, .\slabel{eq:PcbRcb}
\end{subeqnarray}
The leading constants would be unity, $\chi = \theta = 1$, if the radiative zone were replaced by an isothermal layer.  In practice, deviations from unity are modest.    Standard radiative structure calculations (see \App{RCBcorr} for details) give 
\begin{equation}
\label{eq:chi}
\chi \simeq \Big(1-\frac{\delad}{\nabla_{\infty}} \Big)^{-\frac{1}{4-\beta}} \simeq 1.53 \, ,
\end{equation}
for $P\cb \gg P\di$, our regime of interest, and with the radiative temperature gradient at depth, $\nabla_\infty = 1/2$, for our dust opacity.  The radiative zones are thus nearly isothermal, as found by R06.  A numerical integration gives $\theta \simeq 0.556$ for our parameters.   

Given the conditions at the RCB,  the density and temperature profiles along the interior adiabat,
\begin{subeqnarray}
\rho &=& \rho\cb \left[ 1 + {\RB' \over r} - {\RB' \over R\cb}  \right]^{1/(\gamma -1)} \slabel{eq:rhoconv}  \\
T	&=& T\cb \left[ 1 + {\RB' \over r} - {\RB' \over R\cb}  \right] \slabel{eq:Tconv}\, ,
\end{subeqnarray} 
follow from hydrostatic balance.  We introduce an effective Bondi radius
\begin{equation}
\RB' \equiv {G M\co \over C_PT\cb} = {\delad \over \chi} \RB
\end{equation} 
to simplify expressions, with $C_P = \Rg / \delad$ the specific heat capacity at constant pressure.

Deep in the adiabatic interior, where $r \ll R\cb \ll \RB'$, the profiles follow simple power laws,
\begin{subeqnarray}\label{eq:deep}
\rho &\simeq& \rho\cb \left[ {\RB' \over r}\right]^{1/(\gamma -1)} \propto r^{-5/2} \slabel{eq:rhoconvdeep}  \\
T	&\simeq& T\cb {\RB' \over r} = {G M\co \over C_P r} \slabel{eq:Tconvdeep}\, .
\end{subeqnarray} 
While the radial density profile depends on the adiabatic index, the $r^{-1}$ temperature scaling is universal.  In self-gravitating models, the temperature gradient,
\begin{equation} \label{eq:TPsg}
{dT \over dr} = - {G m(r) \over C_P r^2}\, ,
\end{equation} 
gives a profile that is flatter than $T \propto r^{-1}$.

Returning to our non-self-gravitating model, the total specific energy at depth,
\begin{equation}\label{eq:ean}
e = e_g + u = -\delad {GM\co \over r} \, ,
\end{equation} 
is simply proportional to the gravitational potential, $e_g = -GM\co/r$. 

\subsection{Mass, Energy and Luminosity}
\label{MELan}
The most relevant quantities for global cooling are the integrated energy, luminosity, and atmospheric mass.  In our non-self-gravitating limit, the mass of our nearly isothermal radiative zones is less than the convective interior, as shown in \App{iso}.

The atmospheric mass is thus given by the integration of \Eq{eq:rhoconv}, 
\begin{eqnarray} 
\label{eq:Matman}
M_{\rm atm} 
&=& {5 \pi^2 \over 4} \rho\cb {\RB'}^{5/2} \sqrt{R\cb}, 
\end{eqnarray}
in the relevant limit $R\co \ll R\cb \ll \RB'$ and for $\gamma =  7/5$.  Mass is concentrated near the outer regions of the convective zone, a result that holds for $\gamma > 4/3$.  

Using \Eq{eq:PcbRcb} to eliminate $R\cb$, the ratio of atmosphere to core mass becomes
\begin{equation} \label{eq:crit}
{M_{\rm atm} \over M\co} = {P\cb \over \xi P_M}\, ,
\end{equation} 
where we define a characteristic pressure and a logarithmic factor:
\begin{subeqnarray} 
P_M &\equiv& {4 \delad^{3/2} \over 5 \pi^2 \sqrt{\chi} } {G M\co^2 \over {\RB'}^4} \,  \label{eq:PM} \\
\xi &\equiv& \sqrt{\ln[ P\cb/(\theta P_{\rm d})]} \slabel{eq:xidef}\, .
\end{subeqnarray} 

The atmosphere mass increases as radiative losses lower the internal adiabat and increase $P\cb$.  The crossover mass, $M_{\rm atm} = M\co$, is reached when
\begin{equation} \label{eq:Pcbc}
P\cb = \xi P_M ,
\end{equation} 
i.e.\ near the characteristic pressure $P_M$.  The critical value of the order unity factor $\xi$ is found by eliminating $P\cb$ from \Eqs{eq:xidef}{eq:Pcbc}.  This logarithmic factor complicates our analytic description.  Since it remains order unity, we simply hold it fixed in our scalings. 

The total energy is concentrated towards the core if $|e| \rho  r^3 \propto \rho r^2$ drops with increasing $r$.  This condition requires $\gamma < 3/2$, which our choice of $\gamma = 7/5$ satisfies, but $\gamma = 5/3$ (monatomic gas) would not. 

Integration of \Eq{eq:ean} over the mass of the atmosphere thus gives
\begin{subeqnarray} 
E &=& - 4 \pi \nabla_{\rm ad} G M\co \int_{R\co}^{R\cb} \rho r dr \\
&\approx& - 4 \pi P\cb {\RB'}^{1 \over \nabla_{\rm ad}} \left(\gamma-1 \over 3 - 2 \gamma\right)  R\co^{2\gamma-3\over \gamma-1}  \slabel{eq:Ean} \\ 
&\approx& - 8 \pi P\cb {\RB'^{7/2} \over \sqrt{R\co}} \slabel{eq:Eus} \, ,
\end{subeqnarray} 
with $\gamma < 3/2$ and $\gamma = 7/5$ in \Eqs{eq:Ean}{eq:Eus}, respectively. 

The emergent luminosity from the RCB, 
\begin{equation} \label{eq:Lcb}
L\cb = {64 \pi G M\cb \sigma T\cb^4 \over 3 \kappa P\cb } \nabla_{\rm ad} \approx L\di {P_{\rm d} \over P\cb}\, , 
\end{equation} 
follows from \Eq{eq:delrad} and marginal convective stability, $\delrad = \delad$, where we define 
\begin{equation} 
L\di \equiv {64 \pi G M\cb \sigma T_{\rm d}^4 \over 3 \kappa(T_{\rm d}) P_{\rm d}} \nabla_{\rm ad}\chi^{4-\beta}.
\end{equation} 
The scaling $L\cb \propto 1/P\cb$ shows that luminosity drops as the atmosphere cools and $P\cb$ deepens.  This result relies on the pressure independence of  dust opacities.  For fully non-self-gravitating results, we replace $M\cb$, the mass up to the RCB, with the core mass, but the mass of the convective atmosphere can be included for a slightly higher order estimate.

\subsection{Cooling Times \& Core Masses}
\label{coolingan}

Our analytic cooling model uses $L = -\dot{E}$,  neglecting the surface terms in \Eq{eq:coolingglobal}.\footnote{\App{surfterms} shows that these terms are negligible for a non-self-gravitating model.}  
Applying \Eqs{eq:Eus}{eq:Lcb}, the time it takes to cool the atmosphere until the RCB reaches a given pressure depth, $P\cb$, is 
\begin{subeqnarray} 
t_{\rm  cool} &=&- \int_{P\di}^{P\cb } {d E/d P\cb \over L \cb} dP\cb \\
&\approx& 4 \pi {P\cb^{2} \over P\di} {\RB'^{7/2} \over L\di \sqrt{R\co}} \label{eq:tcool}\, .
\end{subeqnarray} 
The initial RCB depth is set to $P\di$ as a formality.  The cooling slows as it proceeds with $t_{\rm cool} \propto P\cb^2$.  

We expect runaway growth to begin around the crossover mass, $M_{\rm{atm}} = M_{\rm c}$. \Eqs{eq:tcool}{eq:Pcbc} give the time to crossover as
\begin{eqnarray} 
\label{eq:tcoolan}
t_{\rm co} &\approx & 2 \times 10^8 {F_T^{5/2}  F_\kappa \left(\xi \over 3.4\right)^2  \over \mcn{10}^{5/3} \aun{10}^{15 / 14}} \yr \, .
\end{eqnarray} 
We estimate the critical core mass, $\MC$, by equating the crossover time with a typical protoplanetary disk lifetime,
\begin{equation}
t\di = 3 \times 10^6 ~{\rm yr} \,.
\end{equation} 
Setting $t_{\rm co} = t\di$ gives
\begin{equation}\label{eq:Mcrit}
\MC \approx 100 {F_T^{3/2} F_\kappa^{3/5}   \left(\xi \over 2.6 \right)^{6/5} \over \aun{10}^{9 / 14}} \; M_\oplus.
\end{equation} 

Both $t_{\rm co}$ and $\MC$ are too large to be interesting, or to be correct based on previous results and the numerical results in this paper.  The main reason for this discrepancy is the neglect of self-gravity.  Section \ref{sec:KH} explores in detail the effects of self-gravity on atmospheric structure, luminosity and evolution.  One might expect self-gravity to be only a modest correction for $M_{\rm atm} \leq M\co$.   This seemingly reasonable expectation can be misleading, an interesting result in itself.

Aside from this insight, the analytic model is useful because, despite the amplitude error, it explains the basic scaling of numerical cooling models.  As is well known \citep{HubBod05}, a lower opacity, here scaling with $F_\kappa$, allows faster cooling and gives a lower critical core mass.  The cooling timescale and critical core mass depend only weakly on the disk pressure, via the logarithmic factor $\xi$.   The almost exponential rise in radiative zone pressure with depth, see \Eq{eq:PcbRcb}, explains why disk pressure has only a weak effect on cooling at the RCB.  

Lower disk temperatures decrease both $t_{\rm co}$ and $\MC$.  The decline in both quantities with semimajor axis is completely explained by the temperature  profile (ignoring the logarithmic factor $\xi$).  The temperature dependence is a competition between two main effects.   The Bondi radius, which quantifies the strength of gravity, decreases for lower temperatures. On the other hand, the luminosity, $\propto T^{4-\beta}$, is lower at colder temperatures, which slows cooling and opposes the overall effect.  With $t_{\rm co} \propto F_T^{ \beta + 1/2}$, the slope of the dust opacity,  $\beta$, is an important factor in regulating the resulting growth time.

To facilitate comparison with our numerical results, we rescale the analytic results as follows.   Since growth times are too slow in the non-self-gravitating analytic model, we modify the runaway growth criterion to occur at an effective crossover mass $M_{\rm atm} = f M\co$, with $f < 1$.  Specifically, we choose $f = 0.13$, because with this value the analytic model gives the same critical core mass at 10 AU as the numerical model, for our standard choices of other parameters.   This prescription does not mean that runaway growth physically occurs at low atmosphere masses, nor does it cause perfect agreement between the models.  Rather by artificially accelerating growth in the analytic model, we can more easily compare the parameter scalings of the two models.  Mathematically, the modified crossover mass amounts to replacing $\xi \rightarrow f\xi$ in our analysis.  With $f = 0.13$, the revised scalings are
\begin{subeqnarray}
t_{\rm run} & \approx & 3 \times 10^6 {F_T^{5/2}  F_\kappa \left(\xi \over 3.4\right)^2  \over \mcn{10}^{5/3} \aun{10}^{15 / 14}} \yr  \slabel{eq:tcoolanf} \\
\MC & \approx & 8\, {F_T^{3/2} F_\kappa^{3/5}   \left(\xi \over 2.6 \right)^{6/5} \over \aun{10}^{9 / 14}} \; M_\oplus\ \, ,  \slabel{eq:Mcritf}
\end{subeqnarray} 
now using $t_{\rm run}$, defined as the time when $M_{\rm atm} = f M\co$,  instead of $t_{\rm co}$ for the runaway growth timescale.  When plotting these rescaled analytic results, we include the variations in $\xi$ from \Eq{eq:xidef}.

\section{Quasi-Static Kelvin-Helmholtz Contraction}
\label{sec:KH}

We examine the structure and evolution of our model atmospheres, calculated as described in \S\ref{sec:model}.  For comparison, we also show results of the non-self-gravitating analytic model of \S\ref{sec:coolingan}.  Radial structure is presented in \S\ref{sec:profiles}, time evolution is described in \S\ref{sec:timeev}, and the validity of our two-layer cooling model is examined in \S\ref{sec:endoftime}.

\begin{figure}[tb]
\centering
\includegraphics[width=0.5\textwidth]{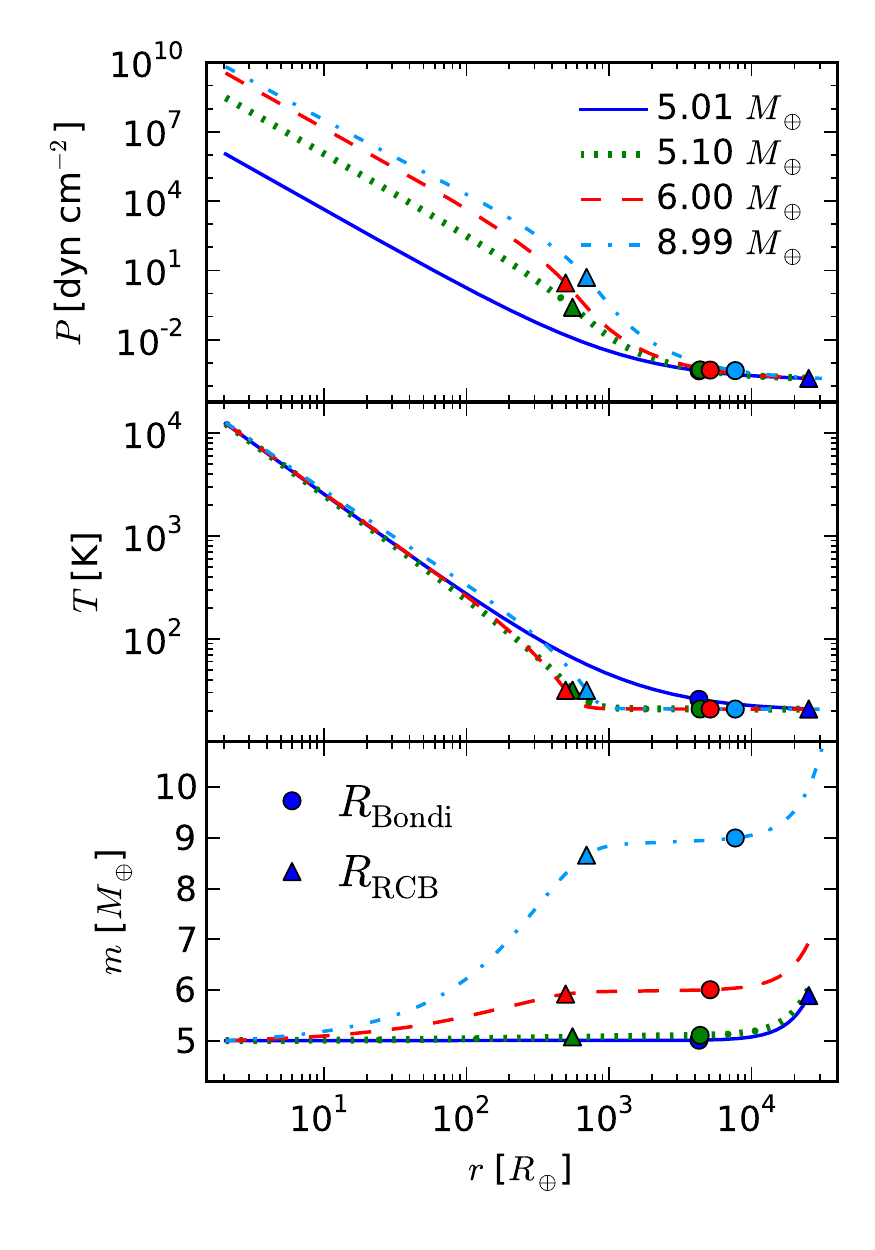}
\caption{Radial profiles of atmospheric pressure, temperature and enclosed mass (core included) for a $5 M_{\oplus}$ core at $60$ AU.   Solid, dotted, dashed and dot-dashed lines correspond to solutions with total mass (core and atmosphere) of $5.01 M_{\oplus}$, $5.10 M_{\oplus}$, $6.00 M_{\oplus}$ and $8.99 M_{\oplus}$, respectively (see text for significance of these masses).  Circles and triangles mark the locations of the Bondi radii and of the radiative-convective boundaries, respectively.  The radial profiles extend from the core to the Hill radius.} 
\label{fig:profiles}
\end{figure}

\subsection{Atmospheric Structure}
\label{sec:profiles}
Figure \ref{fig:profiles} shows radial profiles at different stages of atmospheric growth around a $5 M_{\oplus}$ core at $60$ AU.  Quoted mass values include the core plus atmosphere within the smaller of $\RB$ or $\RH$, which for these cases is $\RB$.  The 8.99 $M_{\oplus}$ solution is the mass 
that satisfies our runaway growth criteria (described in \S\ref{sec:timeev}). 

The lowest mass atmosphere -- which we take as our initial state -- is fully convective and shares the disk's entropy.  In \Fig{fig:profiles} this state is the 5.01 $M_{\oplus}$ solution with no radiative zone.

Cooling and contraction allow the atmosphere to accrete more gas.  In the convective zones, higher mass solutions have lower entropy and higher pressures.  A radiative zone emerges to connect the lower entropy interior to the higher entropy disk.  \Fig{fig:profiles} shows that this radiative zone is already fairly deep in the 5.10 $M_\oplus$ solution. 
 
The atmospheric structure is well approximated by our non-self-gravitating, analytic solutions.  Deep in convective zones, thermal energy is a fixed fraction of the gravitational potential energy, giving $T \propto r^{-1}$ and $P\propto r^{-1/\delad}$  as in \Eq{eq:deep}.  This behavior is seen in \Fig{fig:profiles} for $r\ll R\cb$. Near the core, \Eq{eq:Tconvdeep} gives the core temperature, $T\co= G M\co / (C_{\rm P} R\co)$, unaffected by  the overlying atmosphere mass which does not contribute to the gravitational potential.  Closer to the RCB, self-gravity is no longer negligible, particularly for large envelope masses. Instead, these high mass solutions show a slightly flatter profile in $T$ and also in $P \propto T^{1/\delad}$, as explained by \Eq{eq:TPsg}.

In agreement with \Eq{eq:cb2}, the radiative zones remain nearly isothermal, even for the higher masses.  Consequently, the pressure increases nearly exponentially with depth.

\subsection{Time Evolution}\label{sec:timeev}

\begin{figure}[tb]
\centering
\includegraphics[width=0.5\textwidth]{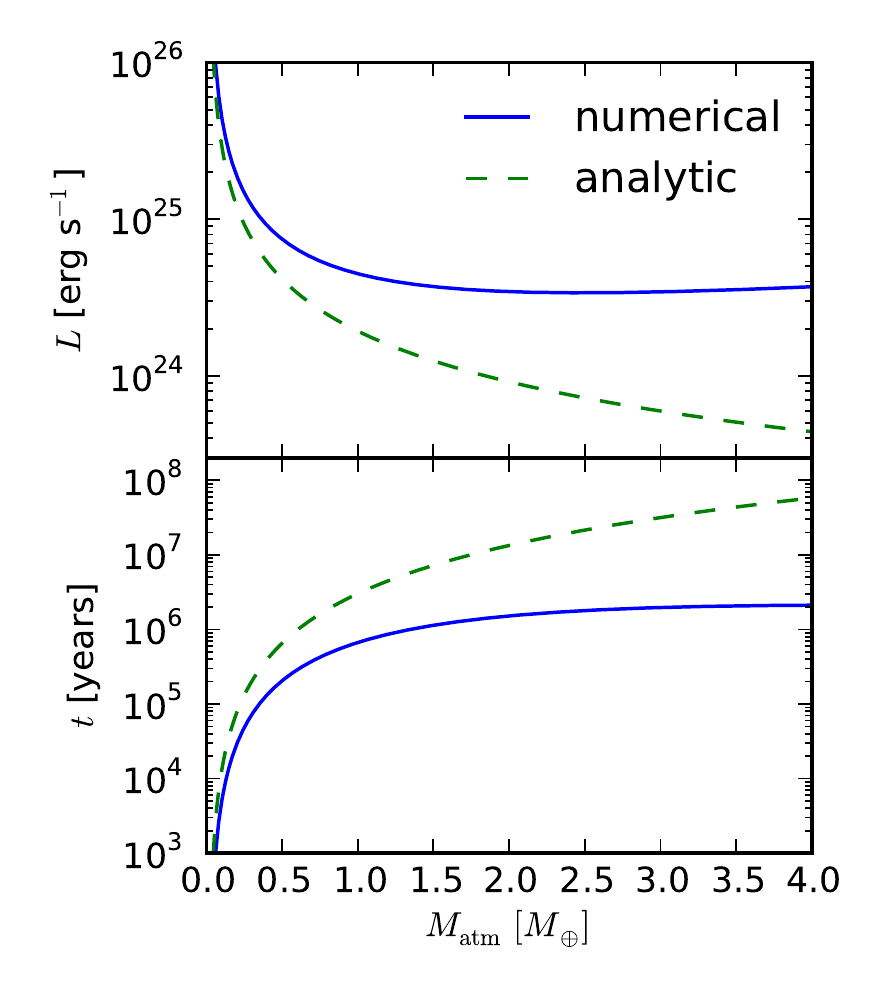}
\caption{Evolution of the luminosity and elapsed time during atmospheric growth around a $5 M_{\oplus}$ core at $60$ AU.  The luminosity is initially high, then decreases as the atmosphere grows in mass and the radiative zone becomes optically thicker.  Due to the neglect of self-gravity, the analytic model (\emph{dashed curve}) gives luminosities that are too low and evolution times that are too long.}
\label{fig:Ltplot}
\end{figure}

The cooling model of \S\ref{sec:twolayer} is used to connect solutions with different atmospheric masses into an evolutionary sequence.  \Fig{fig:Ltplot} shows the luminosity evolution and the elapsed time as a function of atmospheric mass for the same parameters as in \Fig{fig:profiles}.

During the early stages of atmospheric growth, the luminosity drops sharply.  This behavior is seen in both the full numerical solutions and the analytic model.  With increasing atmospheric mass, the pressure depth of the RCB increases, along with the optical depth ($\propto \kappa P\cb)$.  Consequently, the radiative luminosity decreases.  This behavior is described in \Eqs{eq:crit}{eq:Lcb}.

At later stages of evolution, the numerical model in \Fig{fig:Ltplot} shows a flat luminosity with increasing mass and also time (not shown).  By contrast, the non-self-gravitating analytic model gives a luminosity that continues to drop as the atmosphere becomes more massive.  To understand this difference, consider the scaling of \Eq{eq:Lcb}, $L\cb \propto M\cb T\cb^4/ (\kappa\cb  P\cb)$, which holds in both cases.   Accounting for the higher enclosed mass in the self-gravitating model gives a somewhat higher luminosity, as desired.  However, the main effect is that \Eq{eq:crit} -- which describes a nearly linear relation between atmospheric mass and RCB pressure --  breaks down for self-gravitating solutions.  This behavior can be seen in the top panel of \Fig{fig:profiles} where the $P\cb$ increases significantly from 5.10 to 6.0 $M_\oplus$, but only increases relatively modestly with further growth to 8.99 $M_\oplus$.   In the higher mass solutions, the relatively low $P\cb$ values (and thus the relatively high luminosities) require an outward shift in $R\cb$, as shown in \Fig{fig:profiles}. This shift does not occur in the analytic solution, where $R\cb$ continually decreases with atmospheric mass (cf. Equations \ref{eq:PcbRcb} and \ref{eq:crit}).

The accelerated growth in the numerical model, as shown in the bottom panel of \Fig{fig:Ltplot}, is also a direct result of the higher cooling luminosities with self-gravity included.  Even when $M_{\rm atm} / M\co$ is only few percent, i.e.\ well before the crossover mass, the effect of self-gravity is  quite evident.  Closer to the star, the effects of self-gravity are not as strong for low atmosphere masses.  Nevertheless, all our models show that self-gravity noticeably accelerates growth for $M_{\rm atm} > 0.1 M\co$.

\begin{figure}[tb]
\centering
\includegraphics[width=0.5\textwidth]{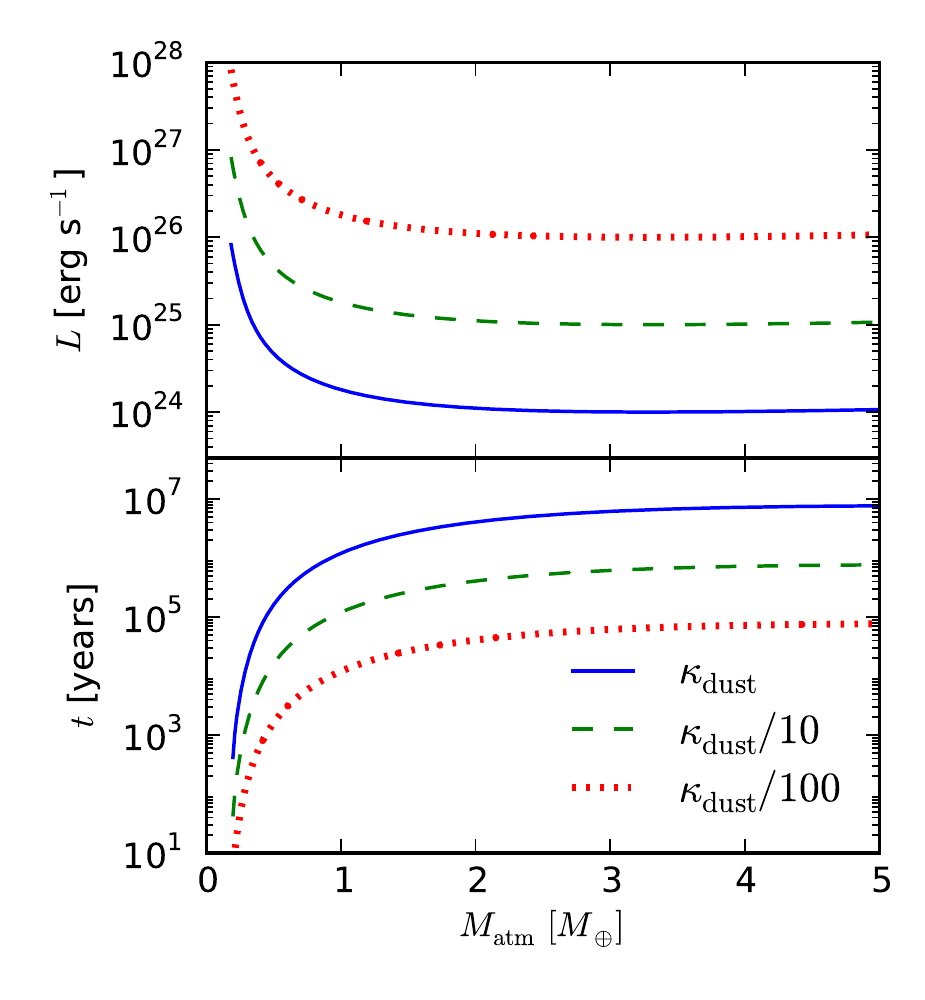}
\caption{The effect of dust abundance on atmospheric evolution.   Reducing dust opacities by factors of 10 and 100 from standard Solar abundances gives higher luminosities and faster atmospheric growth.  Plotted quantities are similar to \Fig{fig:Ltplot}, but for a $5 M_{\oplus}$ core at 10 AU. }  
\label{fig:LtvsMopacity}
\end{figure}

In \Fig{fig:LtvsMopacity}, our standard dust opacity, \Eq{eq:opacitylaw}, is reduced by factors of 10 and 100.  Lower opacities result in higher luminosities and faster evolution.  Our model thus confirms a well-established result \citep{HubBod05}.  While clearly an important effect, atmospheric dust opacities are difficult to robustly predict. Ablation of infalling solids is a dust source.  Sinks include the sequestration of solids in the core and dust settling through the radiative zone.  Grain growth both reduces dust opacities per unit mass and favors settling.  Our scenario of negligible ongoing particle accretion tends to favor low dust opacities.   To be conservative, however, our reference case considers full Solar abundances. The effect of opacity reduction on the critical core mass is described in \S\ref{sec:critical}. 

\begin{figure}[tb]
\centering
\includegraphics[width=0.5\textwidth]{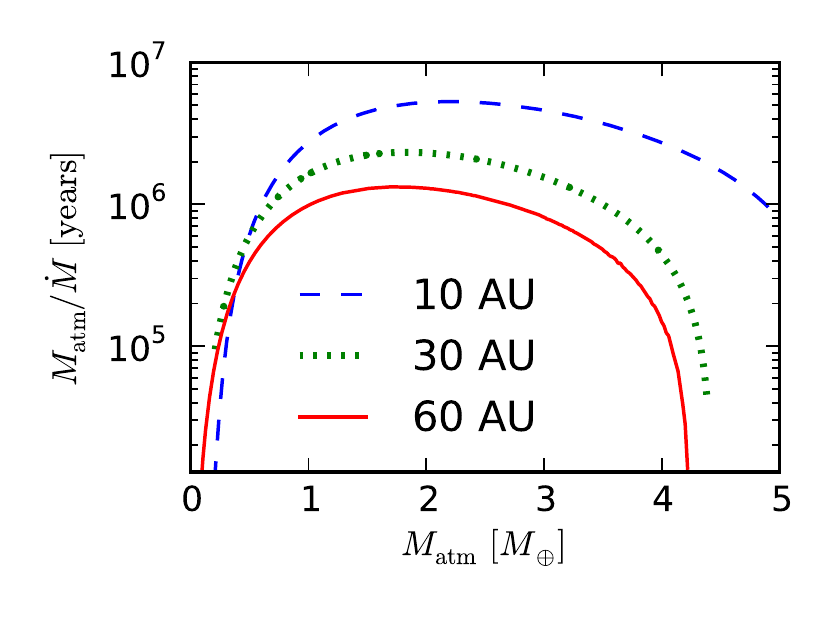}
\caption{Evolution of the atmospheric growth timescale with mass around a $5 M_{\oplus}$ solid core  located at 10, 30 or 60 AU, for standard Solar opacities.  Growth is slowest for $M_{\mathrm{atm}} \sim 1 - 3 M_{\oplus}$, i.e.\ before the crossover mass at $M_{\rm atm} = M\co$.}
\label{fig:growthtime}
\end{figure}

\Fig{fig:growthtime} plots the evolution of the atmospheric growth timescale, $M_{\rm atm}/\dot{M}$,  around a $5 M_{\oplus}$ core at several locations in our reference disk model.  This instantaneous growth time shows clearly that the atmosphere spends the bulk of its time growing though intermediate atmospheric masses, $\sim 1 -3$ $M_\oplus$ in this case.  Growth times are short both early -- when the radiative zone is transparent -- and late -- when self-gravity accelerates growth.  

The fact that growth times have a well defined maximum is a characteristic of accelerating growth.  Unlike our analytic model, which must assume that runaway growth begins near the crossover mass, our numerical model allows us to measure when runaway accretion starts.  Runaway growth does not begin at a universal value of $M_{\rm atm}/M\co$.  Further from the star, runaway growth begins at smaller $M_{\rm atm}/M\co$, as \Fig{fig:growthtime} shows.  \Fig{fig:tvsM} (described in the next section) shows how the onset of runaway growth depends on core mass. 

We quantify the runaway growth timescale, $t_{\rm run}$, as the time when $M_{\rm atm}/\dot{M}$ drops to 10\% of its maximum value.  The choice of 10\% is arbitrary; the precise threshold chosen is relatively unimportant because growth continues to accelerate.


\subsection{Validity of the Two-Layer Cooling Model}
\label{sec:endoftime}

We examine the  validity of our cooling model by comparing our model luminosity to the neglected luminosity, $L_{\rm negl}$,  that a more detailed model would generate in the radiative zone.  We compute $L_{\rm negl}$ from the entropy difference between successive radiative zone solutions.  We then integrate the energy equation, $\p L / \p m = - T \p S/ \p t$, over the average depth of the radiative zone.\footnote{While useful as a diagnostic, the neglected luminosity cannot reliably correct the global cooling model because the effects of $L_{\rm negl}$ on the structure of the radiative zone are still ignored.}

\begin{figure}[tb]
\centering
\includegraphics[width=0.5\textwidth]{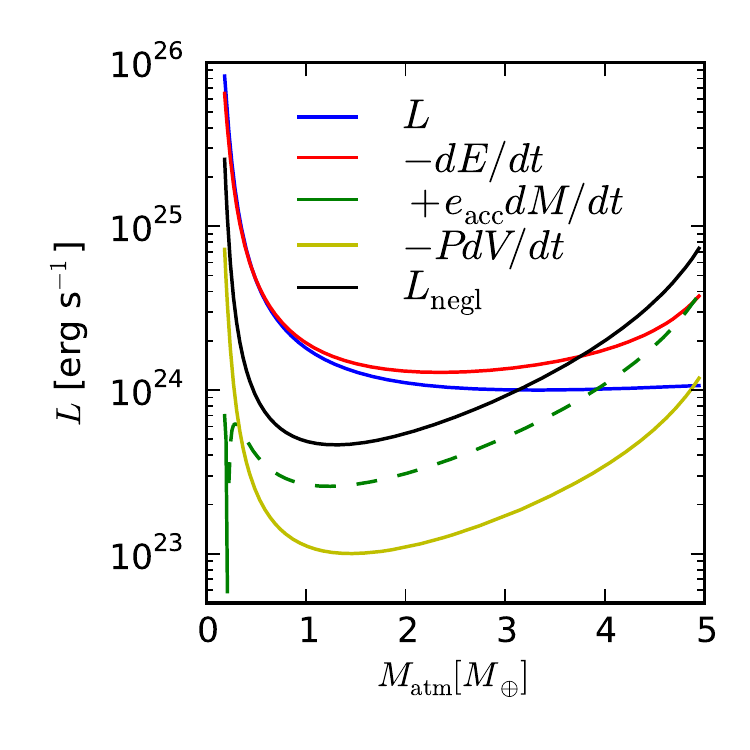}
\caption{Individual terms in the atmospheric cooling model of \Eq{eq:coolingglobal}, for a $5 M_\oplus$ core at 10 AU.  The dashed curve for accretion energy indicates a negative contribution.  All quantities are evaluated at the RCB, except for $L_{\rm negl}$, the extra luminosity that would have been generated in the radiative zone, but is neglected in our model. The neglected luminosity is a small correction to the model luminosity $L$ for $M_{\rm atm} \lesssim 3 M_{\oplus}$.   Since these low masses dominate growth times, our model is roughly accurate.}
\label{fig:coolingterms}
\end{figure}

 \Fig{fig:coolingterms} shows that the neglected luminosity is indeed negligible during the early stages of evolution.  However, $L_{\rm negl}$ exceeds the model luminosity, $L$, at high masses, $M_{\rm atm} > 3 M_\oplus$ in this case.  Our cooling model is thus inaccurate at higher masses.  However, the model remains reasonably accurate up to the beginning stages of runaway growth, which is sufficient for our purposes of widely exploring parameter space and exploring trends.
 
The individual terms in the global cooling model of \Eq{eq:coolingglobal}, evaluated at the RCB, are also plotted in \Fig{fig:coolingterms}.  At low masses, the change in energy, $- \dot{E}$, makes the dominant contribution to luminosity.  As the mass increases, the surface terms become more significant, led by the accretion energy.  However, the surface terms are everywhere smaller than $L_{\rm negl}$.  Thus wherever our model is accurate -- including the crucial early phases of growth -- surface terms are a minor correction.  The neglect of surface terms in the analytic model is thus not a serious omission.


\section{Results for Giant Planet Formation}
\label{sec:critical}

We now use our structure and evolution models to estimate the timescales and minimum core masses for giant planet formation for a range of disk conditions and other model parameters.  Our results for atmosphere growth times -- the time for a core of fixed mass to undergo runaway gas accretion -- are presented in \S\ref{sec:tcross}.  Section \ref{sec:critcore} gives our results for critical core masses, the minimum values that trigger runaway atmospheric growth within a plausible disk lifetime, here 3 Myr.

Our models focus on giant planet formation between 5 and 100 AU, as the outer disk is of particular interest for direct imaging searches.   The growth of atmospheres close to the star is also important, but spherical accretion models (including ours) are less applicable here.  In the inner disk, critical core masses increase, yet lower mass planets start to open gaps and outgrow the disk scale height, see \Eq{eq:Mth}.  These concerns prevent us from applying our model to the inner disk.

\subsection{Runaway Growth Timescale}
\label{sec:tcross}
The time to undergo atmospheric runaway growth, $t_{\rm run}$, sets a minimum timescale for the formation of giant planets.  Due to the accelerating nature of runaway growth, the precise threshold chosen for $t_{\rm run}$ (explained in \S\ref{sec:timeev}) is of minor significance.

\begin{figure}[tb]
\hspace{-.1in}
\includegraphics[width=0.5\textwidth]{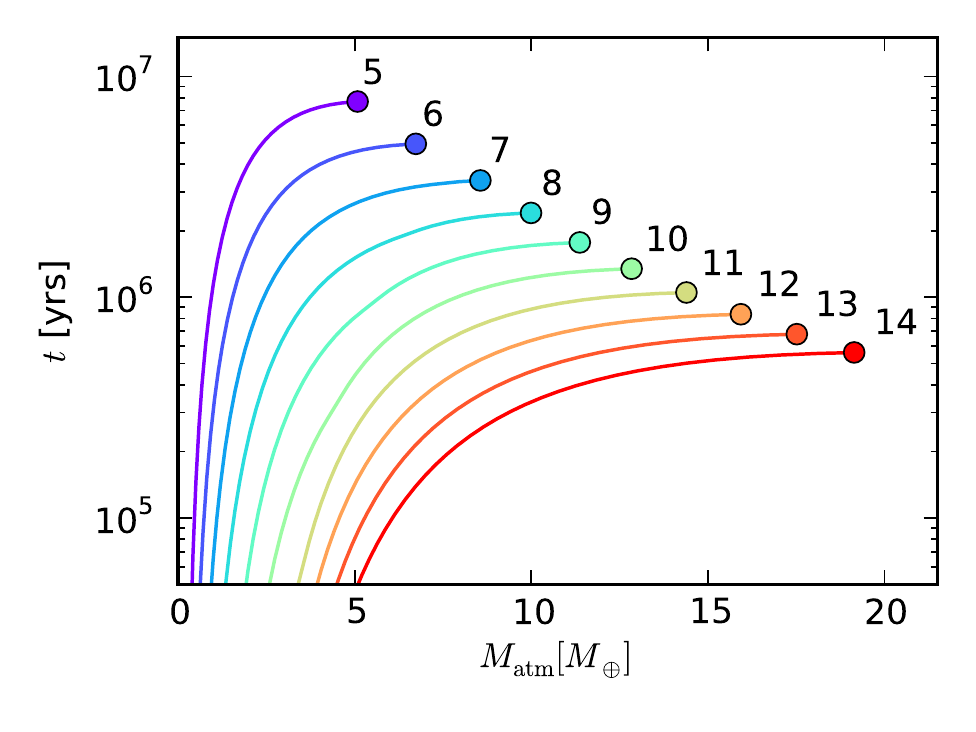}
\caption{Time to grow an atmosphere of mass $M_{\rm{atm}}$ for cores with fixed masses between $5 M_{\oplus}$ and $14 M_{\oplus}$ (as labeled) at $10$ AU in our fiducial disk. Circles mark the runaway growth time, $t_{\rm run}$, which occurs at roughly the crossover mass, $M_{\rm{atm}} = M\co$.  Both the time to reach a fixed atmosphere mass and the runaway growth time are shorter for larger cores. For larger $M\co$, runaway growth commences at higher $M_{\rm atm}/M\co$ values.}
\label{fig:tvsM}
\end{figure}

\begin{figure}[htb]
\centering
\includegraphics[width=0.48\textwidth]{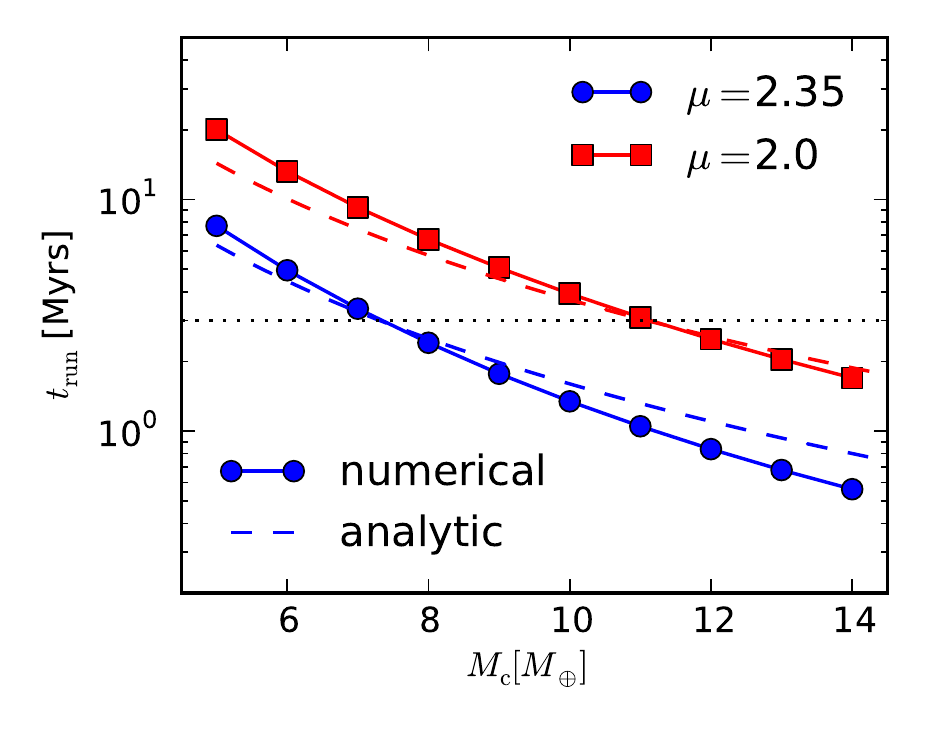}
\caption{Runaway growth time, $t_{\rm{run}}$, vs.\ core mass at $10$ AU, for two values of the mean molecular weight.  Our numerical model (\emph{solid curves}) is compared to our non-self-gravitating analytic model (\emph{dashed curves}, from Equation \ref{eq:tcoolanf}).  A typical protoplanetary disk life time of $3$ Myr is plotted for comparison. The runaway growth time is larger for a lower mean molecular weight.}
\label{fig:tvsMcomp}
\end{figure}

\subsubsection{Effects of Core Mass}
\label{Mct}

\Fig{fig:tvsM} shows the growth of atmospheric mass with time for several core masses at $10$ AU in our fiducial disk.  Atmospheres grow faster around more massive cores due to stronger gravitational binding.  The endpoint of each curve marks $t_{\rm run}$.   Runaway growth occurs near the crossover mass, when $M_{\rm atm} \sim M\co$, in agreement with previous studies. Lower core masses undergo runaway accretion at fractionally smaller atmosphere masses. 

\Fig{fig:tvsMcomp} shows how $t_{\rm run}$ varies with core mass, also at 10 AU.  The numerical results are plotted against our non-self-gravitating analytic model, described in \S\ref{sec:coolingan}.  The analytic model reproduces the general decline in $t_{\rm run}$ with core mass.  The numerical model, which includes self-gravity, has a somewhat steeper mass dependence.  A modest correction due to self-gravity is unsurprising, and consistent with the above-mentioned trend in $M_{\rm atm}/M\co$ ratios.  Moreover, in the analytic theory, crucial quantities like $M_{\rm atm}$ and $L$ (both roughly $\propto M\co^3$ near crossover) have non-linear dependence on core mass, offering plenty of opportunity for self-gravitational corrections.

The effect of mean molecular weight is also shown in \Fig{fig:tvsMcomp}.  A lower $\mu$ gives longer growth times, because more cooling is required to compress the atmosphere.  This effect is both well established in core accretion studies \citep{stevenson82} and intuitive since the atmospheric scale height $\propto 1/\mu$ is more extended for lower $\mu$.   Moreover, the Bondi radius decreases as $\RB \propto \mu$, giving a smaller gravitational sphere of influence (and a weaker compression at $\RH$ when that is the more relevant scale).  To see how this affects cooling times, note that the characteristic RCB depth near runaway scales as $P\cb \sim P_{\rm M} \propto \mu^{-4}$ from \Eq{eq:PM}.  Thus $L \propto 1/P\cb \propto \mu^4$ explains the trend of slower cooling for lower $\mu$.

For $\mu = 2.0$, which represents the idealized case of an H$_2$ atmosphere completely devoid of Helium, $t_{\rm run}$ increases by factors of $\sim 2 - 3$.  Thus fairly drastic changes in atmospheric composition are required for $\mu$ to significantly affect core accretion timescales. In principle, changes in the mean molecular weight of the gas also affect the EOS, including $\delad$, but such effects are not considered here (see \S\ref{sec:EOS}).

\begin{figure*}[tb]
\centering
\includegraphics[width=1.\textwidth]{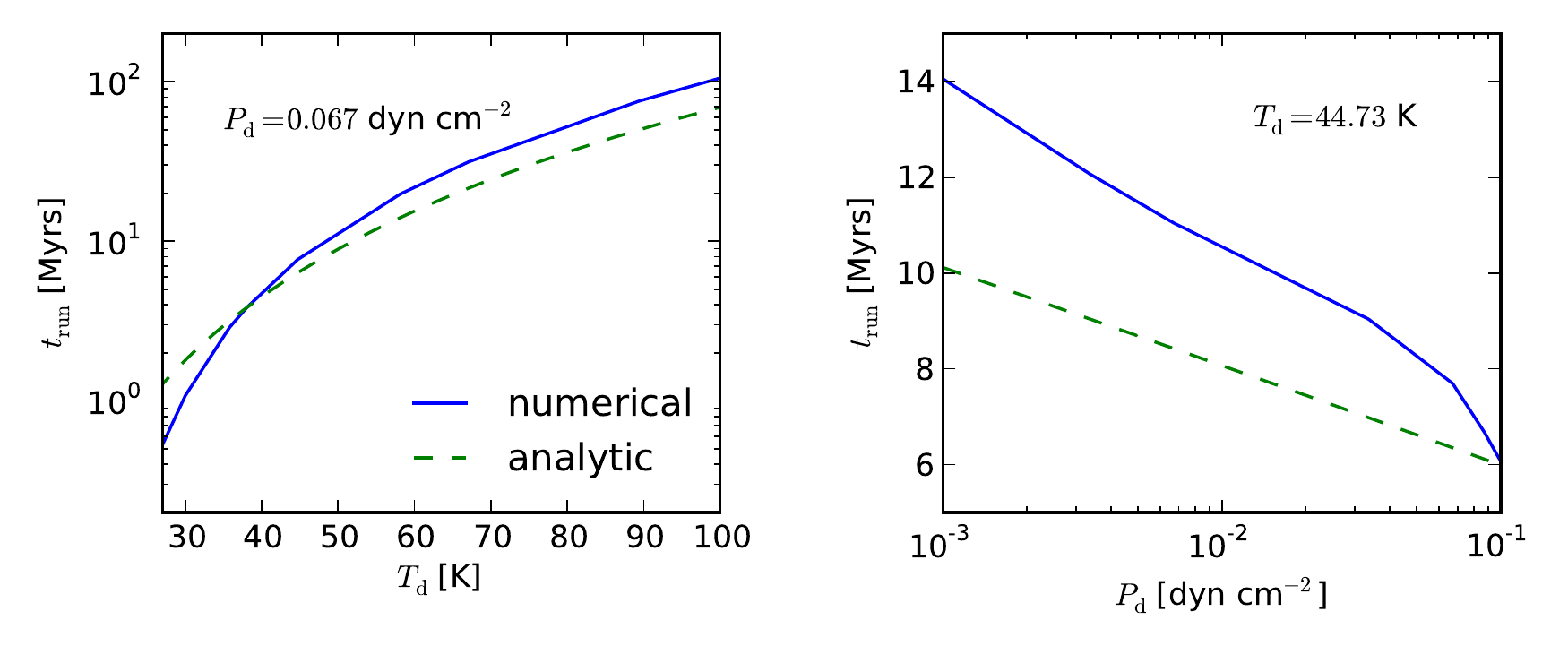}
\vspace{-0.3in}
\caption{Runaway growth time as a function of disk temperature (\emph{left}) and pressure (\emph{right}) around a $M_{\rm c} = 5 M_{\oplus}$ core.   The disk pressure or temperature (\emph{left} or \emph{right}, respectively) are fixed at values for 10 AU in our disk model.   The analytic scalings given by \Eq{eq:tcoolanf} are plotted for comparison, as described in the text.  Gas accretion slows down significantly at higher temperatures, but only speeds up modestly as the disk pressure or density increase.} 
\label{fig:TPeffects}
\end{figure*}

\subsubsection{Effects of Disk Temperature and Pressure}
\label{sec:TPeffects}

\Fig{fig:TPeffects} shows how the runaway accretion time varies with disk temperature, $T\di$, or pressure, $P\di$, holding the other quantity fixed.   The analytic model roughly reproduces the temperature and pressure scalings, again with some discrepancies due mainly to the neglect of self-gravity.  Temperature variations are much more significant than pressure variations (note the difference in logarithmic and linear axes).  Since midplane disk conditions depend only on temperature and pressure in our model,\footnote{See \Eq{eq:diskparam}.  When gap opening is considered in models of later growth stages, the orbital frequency and effective viscosity become relevant as well.} the dominant effect of disk location is temperature.  

The decline in growth times with lower temperatures arises from a balance of competing effects.  The cooling luminosity is inherently smaller at lower temperatures.  Overpowering this effect, the larger Bondi radius and lower dust opacity act to accelerate growth at lower temperatures.

Growth times depend only weakly on, but do fall slightly with, pressure.   This result may be surprising, given that the disk is the source of atmospheric mass and the atmosphere must match onto the disk's density and pressure.  The nearly exponential increase in pressure with depth through the radiative zone explains this effect.  Cooling is largely regulated at the RCB, and a modest change in RCB depth compensates for large variations in disk pressure.

Section \ref{sec:coolingan} shows how these temperature and pressure effects arise in our analytic model.

\begin{figure}[htb]
\centering
\includegraphics[width=0.5\textwidth]{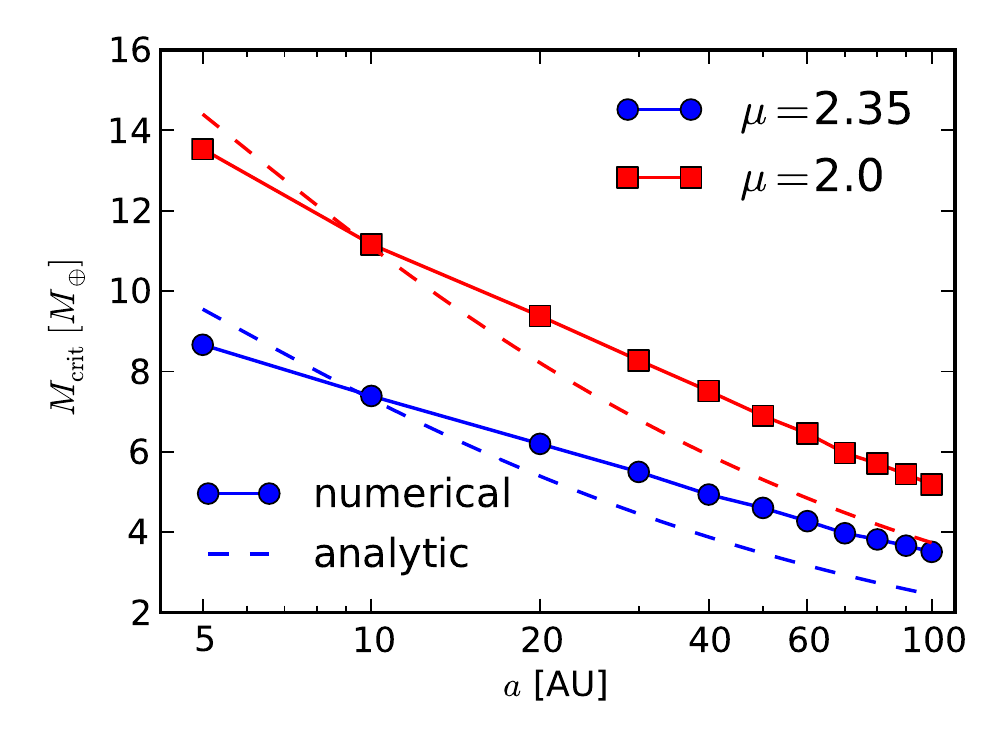}
\caption{The critical core mass as a function of semimajor axis, for a disk lifetime of $3$ Myrs and two values of the mean molecular weight ($\mu = 2.35$ is for Solar abundances). The decline in $\MC$ with distance is a robust result for standard disk models.  The analytic model, which neglects self-gravity, over-predicts the steepness of the decline.}
\label{fig:Mcvsa}
\end{figure}

\begin{figure}[htb]
\centering
\includegraphics[width=0.5\textwidth]{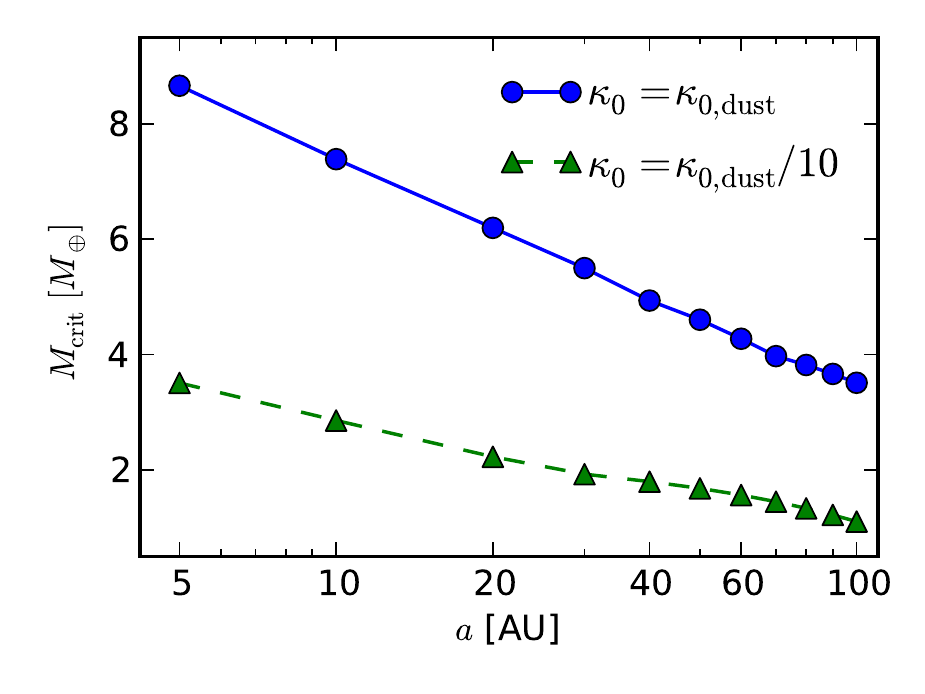}
\caption{Critical core masses vs. distance for standard and reduced (by a factor of ten) dust opacities.  Lower opacities give significantly lower $\MC$ values.  Atmospheric opacity remains a large uncertainty in core accretion models.}
\label{fig:Mcritopacity}
\end{figure}

\subsection{Critical Core Mass}
\label{sec:critcore}

The critical core mass declines with distance from the star, as shown in \Fig{fig:Mcvsa} for our standard disk model.   The main reason for the decline, as explained above, is that atmospheres grow faster at lower temperatures.  The lower densities and pressures in the outer disk have a much smaller effect.  Since the vast majority of disk models have a disk temperature that declines with $a$, our qualitative result is robust.   We also show that higher $\mu$ values give lower values of $\MC$.

The average power-law decline from 1 to 100 AU (not plotted) is $\MC \propto a^{-0.3}$, for both choices of $\mu$.   While not a drastic decline, the ability of distant low mass cores to accrete gas efficiently is significant for the interpretation of direct imaging surveys.  However, our model offers no guarantee of copious giant planets at large distances.  Many histories of solid accretion are possible, and solid cores may grow too slowly to allow the rapid gas accretion that we model.  

Our non-self-gravitating analytic model (dashed curves in \Fig{fig:Mcvsa}) over-predicts the steepness of the decline in $\MC$ with $a$.  This discrepancy is not surprising, as we have shown that self-gravity strongly affects evolution, even before runaway growth begins and the crossover mass is reached.

\Fig{fig:Mcritopacity} shows that reducing the opacity by an order of magnitude significantly reduces $\MC$.  Furthermore, this opacity effect is stronger at larger distances.  The opacity reduction by a factor of 10 lowers $\MC$ at 5 AU by a factor of $\sim$$2.5$ and at 100 AU by a factor of $\sim$$3.5$.  Atmospheric opacity is a dominant uncertainty in core accretion modeling.  However, unless atmospheric opacity varies significantly with disk radius, the general decline in $\MC$ with increasing $a$ should hold.  For our scenario of negligible ongoing planetesimal accretion, it is tempting to think that dust could settle out of the radiative zone, lowering the opacity \citep{podolak03}.  Nonetheless, since opacity near the RCB is a crucial factor, we speculate that convective overshoot may prevent very low opacities.

A larger disk lifetime further reduces $M_{\rm crit}$. Recent studies have shown that gas disks may live up to $\propto$$10-12$ Myr \citep{bell13}. As $M_{\rm crit} \sim t\di^{-3/5}$ (cf. Equation \ref{eq:tcoolan}), a disk lifetime of 10 Myrs decreases the critical core mass by a factor of two.  Of course, some fraction of the disk lifetime must be allocated to the growth and possibly migration of the core.

\subsection{Comparison with Previous Studies}
\label{sec:comp}

We can directly compare our results with other studies of protoplanetary atmospheric growth in the absence of planetesimal accretion, notably those in I00 and PN05.  A major distinguishing feature of our study is that we explore a range of disk conditions, while I00 and PN05 focus on growth at the location of Jupiter, 5.2 AU, in their disk models.  Moreover, both I00 and PN05 solve the full time dependent energy equation and consider more detailed EOSs, in addition to other detailed differences in model parameters.  Despite these differences, the qualitative and quantitative agreement with our study is good.

I00 give an approximate analytic fit to their models' envelope formation time (the equivalent of the runaway accretion time $t_{\rm run}$ in our models), at $a=5.2$ AU, 
\begin{equation}
\label{eq:tauenv}
\tau_{\rm env} \sim 3 \times 10^8 \Big(\frac{M\co}{M_{\oplus}}\Big)^{-2.5} \Big(\frac{\kappa}{1 \text{cm$^2$ s$^{-1}$}}\Big) \,\,\, \text{yr}\, .
\end{equation}
Our model reproduces the linear opacity dependence (see \S\ref{sec:timeev}), even though we use a temperature dependent opacity, \Eq{eq:opacitylaw}, instead of a constant $\kappa$.  Our growth times at 5 AU,
\begin{equation}
\label{eq:trun5au}
t_{\rm run} \sim 5 \times 10^8 \Big(\frac{M\co}{M_{\oplus}}\Big)^{-2.4} \,\,\, \text{yr},
\end{equation} 
agree well with the I00 results, both in magnitude and scaling with core mass.

One goal of our study was to explain the extended luminosity minimum in evolutionary models that characterizes phase 2 in core accretion models, as described in the introduction.  Previous work (see Figure 2 in I00 and Figures 3 and 4 in PN05) shows that this luminosity minimum is an intrinsic feature of atmospheric cooling even in the absence of planetesimal accretion.  We not only reproduce this effect in our simplified model (see \Figs{fig:Ltplot}{fig:LtvsMopacity}), but we also show that self-gravity is the essential ingredient to produce a broad luminosity minimum well before the crossover mass.  

Time-dependent studies that incorporate planetesimal accretion generally find larger formation timescales and critical core masses than those of this work.  This result is expected since additional energy from planetesimal accretion limits the ability of the atmosphere to cool. \citet{pollack96} find an evolutionary time and crossover mass of $\sim$$7.5$ Myrs and $\sim$$16 M_{\oplus}$, respectively, for a MMSN disk model at 5.2 AU and interstellar grain opacity. These values, however, decrease to $\sim$$3$ Myrs and $\sim$$12 M_{\oplus}$ if planetesimal accretion is entirely shut off during the gas accretion phase. The core accretion model of \citet{HubBod05} for Jupiter's formation predicts an evolutionary time of 3.3 Myrs for a $10 M_{\oplus}$ core, for an interstellar opacity, which is also consistent with our result at 5 AU.

A direct comparison with studies that only include planetesimal accretion, and neglect atmospheric cooling, is difficult.  Note, however, that I00 establishes the correspondence between the minimum luminosity in evolutionary models and the minimum planetesimal accretion rate needed in static models, which gives rise to the classical critical core mass of static models \citep{mizuno78, stevenson82}.

In terms of modern static studies, our results complement (and borrow some tools from) R06 and \citet{Raf11}.  These studies consider the disk radius dependence of core accretion for protoplanets that continuously accrete planetesimals.  We show that the limits on core accretion at large radial distance claimed by \cite{Raf11} (and references therein) can be overcome if planetesimal accretion shuts off and the atmosphere is allowed to cool.  Nevertheless, the plausibility of rapid core growth followed by negligible subsequent solid accretion admittedly remains uncertain, as discussed in more detail in \S\ref{sec:conclusions}.

\section{Neglected Effects}\label{sec:neglected}
Since one goal of this paper was to obtain a detailed understanding of atmospheric evolution with a simple model, we have necessarily ignored   effects of potential significance.  We briefly address the most important of these and note that a followup work, Piso, Youdin \& Murray-Clay (in prep., hereafter Paper II), will extend our current models to address some of these effects.

\subsection{Hydrodynamic Effects}\label{sec:hydro}
The neglect of hydrodynamical effects in our model is best discussed in terms of the thermal mass, $M_{\rm th}$, and the length scales introduced in \S\ref{sec:scales}.  In the low mass regime, $M\pla < M_{\rm th}/ \sqrt{3}$, where $\RB < \RH$, we assume that hydrostatic balance holds out to the outer boundary at $\RH$.   In this low mass regime, \citet{Orm13} calculated the 2D (radial and azimuthal) flow patterns driven by stellar tides and disk headwinds.     On scales $\gtrsim \RB$ the flows no longer circulate the planet: they belong to the disk.  While the density structure still appears roughly spherical and hydrostatic, these flows could affect the planet's cooling.  We expect such effects to be weak, as heat losses at greater depths dominate planetary cooling, but more study is needed, especially in 3D.


At higher masses, non-hydrostatic effects become more severe.  At $M\pla \gtrsim M_{\rm th}$ planets can open significant gaps \citep{zhu13}.  At yet higher masses accretion instabilities could occur \citep{AylBat12}.  However, in this high mass regime, the spherically symmetric approximation has already broken down. 

Thus by restricting our attention to low masses, neglected hydrodynamic effects should be minor.   Moreover, since $M_{\rm th} \propto a^{6/7}$ increases with disk radius, spherical hydrostatic models like ours have a greater range of applicability in the outer regions of disks.

\subsection{Realistic Opacities}\label{sec:op}

The importance of envelope opacity is an established factor in core accretion calculations (e.g., \citealt{stevenson82}, \citealt{ikoma00}, R06), and several studies explore the influence of opacity on the timescales for giant planet formation in detail (e.g., \citealt{HubBod05}). Treating dust (and total) opacities as a power law in temperature is a simplification.  Opacities drop by order unity when ice grains sublimate for $T \gtrsim 150$ K and they drop by orders of magnitude when silicate grains evaporate above $T \gtrsim 1500$ K \citep{semenov03, FerAle05}.  Grain growth and composition also affect opacities. Our scenario of no ongoing accretion of solids may result in a grain-free atmosphere, which could substantially reduce the critical core mass ($\sim$$1M_{\oplus}$ in the case of Jupiter, \citealt{hori10}). Paper II explores more realistic opacity laws. 

For this work, we justify a simplified dust opacity by the cool temperatures of both the outer disk and our nearly isothermal radiative zones (see \S\ref{sec:profiles}).  While  convective interiors get significantly hotter than 1500 K, opacity does not affect the structure of adiabatic convecting regions.   A possible caveat is the existence of  radiative zones sandwiched inside the convection interior.  Such ``radiative windows" arise if the opacity drop from ice and metal grain sublimation occurs at sufficiently low pressures.  Our two-layer model ignores this possibility, but radiative windows are known to exist in hot Jupiter models \citep{burrows97, ab06} and have been seen in some core accretion models (Lissauer, personal communication).  The role of radiative windows in core accretion models remains to be explored in detail.

\subsection{Equation of State}
\label{sec:EOS}
 
Our model uses an ideal gas law and a polytropic EOS, given by \Eq{eq:idealEOS}.  However, non-ideal effects can affect atmospheric structure and evolution.  In the lower atmosphere, H$_2$ can partially dissociate at high temperatures.  In the upper atmosphere, H$_2$ rotational levels can become depopulated at lower temeratures.   Paper II uses the \citet{saumon95} EOS, with extensions to lower pressures and temperatures as needed, to explore these effects.

\subsection{Core growth}
\label{sec:placc}

Our model purposefully neglects the accretion of solids to study the fastest rates of gas accretion.  Our $\MC$ values thus differ from the $\MC$ values in static models, such as R06.  For similar parameters, we generally obtain lower $\MC$ values than R06 because our atmospheres are not heated by ongoing planetesimal accretion.  Paper II presents a quantitative comparison.



At low planetesimal accretion rates, $\dot{M}\co$, a static model could formally give lower $\MC$ values than our evolutionary calculations.  R06 shows that $\MC \propto \dot{M}\co^{3/5}$ in static models.  The underlying assumption of static models, a negligibly short KH contraction time, fails whenever static models give a lower $\MC$.  Thus our results give a firm lower limit on $\MC$ which complement the results of static models with planetesimal accretion.

\section{Summary} \label{sec:conclusions}

We study the formation of giant planets by the core accretion mechanism.  Our models start with a solid core that is embedded in a gas disk and no longer accreting solids.  We determine -- as a function of disk location, core mass, and the atmosphere's mean molecular weight and opacity -- whether runaway atmospheric growth can occur within a typical disk lifetime of 3 Myr.  By neglecting the accretion luminosity of planetesimals and smaller solids, we obtain the fastest allowed rate of gas accretion.  

We address core accretion in the outer disk, as it is relevant to direct imaging surveys.  Our model approximations, including spherical accretion and low mass radiative exteriors with opacities dominated by dust, are tuned to conditions in the outer disk.   Our main findings are as follows:

\begin{enumerate}
\item The minimum or critical core mass, $\MC$, for giant planet formation declines with stellocentric distance in standard protoplanetary disk models.  For our reference case, the critical mass is $\sim$$8.5 M_{\oplus}$ at $a = 5$ AU, decreasing to $\sim$$3.5 M_{\oplus}$ at $a =100$ AU.  This decline roughly follows $\MC \propto a^{-0.3}$.

\item The drop in disk temperature with radial distance explains the decrease in critical core masses.  The lower pressures and densities in the outer disk only weakly suppress atmospheric growth.

\item Reducing dust opacities by a factor of 10 reduces critical core masses by a factor of $\sim$$3$.  This reduction is somewhat stronger (weaker) at larger (smaller) separations from the star.

\item A larger mean molecular weight reduces critical core masses, in agreement with \citet{HorIko11}.  If enrichment in heavy elements correlates with increased dust opacity, then the stronger opacity effect will dominate, increasing $\MC$.

\item Runaway growth begins roughly at the crossover mass, when atmosphere and core masses are equal, $M_{\rm atm} \sim M\co$, in agreement with previous work \citep{pollack96}.  Further from the star, runaway growth begins at smaller $M_{\rm atm} / M\co$ ratios.  For larger core masses, runaway growth begins at larger values of $M_{\rm atm} / M\co$.

\item Self-gravity affects atmospheric evolution before crossover.  Significant self-gravitational corrections appear when the atmosphere is only $\sim$$10 \%$ as massive as the core.

\end{enumerate}

Rapid gas accretion onto low mass cores could explain the origin of distant directly imaged giant planets \citep{marois08, lagrange10}.  However, our model does not address the details of how solid cores grow, as many possibilities exist and many uncertainties remain. 
For a giant planet to form with a core near the minimum masses we derive, core growth must  first be rapid and then slow significantly, as in phases 1 and 2, respectively, of \cite{pollack96}.

Initial core growth must be fast, compared to the disk lifetime, to get a sufficiently massive core.  Such rapid core growth is possible in a variety of scenarios, including the fastest gas-free planetesimal accretion rates \citep{dones93} and -- probably more relevantly for gas rich disks --  the aerodynamic accretion of mm-m sized ``pebbles" and ``boulders" in gas disks \citep{ormel10, lambrechts12}.  Moreover, cores could form rapidly closer to the star, then migrate or be scattered outwards by already formed giants \citep{ida13}. 
 
A stronger constraint is that core growth subsequently slow severely, to allow the atmosphere to cool and contract.  To be more quantitative, the minimum cooling luminosity in \Fig{fig:Ltplot}, $L\approx 3.5 \times 10^{24}\;\mathrm{erg s}^{-1}$, could be cancelled by low levels of heating from solid accretion.   A core mass doubling timescale of $\sim$$400$ Myr, or faster, would thus provide enough heating to stall atmospheric cooling and growth.   An additional concern is that isolation masses tend to grow with disk radius, as $M_{\rm iso} \propto \varSigma_{\rm p}^{3/2} a^2 \propto a^{3/4}$ under the approximation that the surface density of accreted planetesimals ($\varSigma_{\rm p}$) scales with the gas \citep{youdin13}.   While this behavior is nominally inconsistent with final core masses that decline with distance, the predictive power of the isolation mass is imperfect.   For starters, the efficiency of planetesimal formation remains uncertain.  Moreover, the locality of core growth, which underlies the isolation mass, disappears when accreted solids drift and/or cores migrate significantly. 
  
Thus while our calculations show that low mass cores can grow into gas giants in the outer disk, ongoing solid accretion could prevent significant atmospheric growth.  In the Solar System, the ice giants Uranus and Neptune,  with core (here ice and rock) masses of $\sim 13 -15$ M$_\oplus$, argue for the latter possibility.  Ongoing exoplanet imaging surveys and their successors \citep{hinz12, macintosh12, close14} will help discriminate among the various planet formation pathways in the outskirts of protoplanetary disks.

\vspace*{5 mm}

\acknowledgements{We thank the referee for a thorough report and Ruth Murray-Clay for detailed feedback and skillful proof reading. We appreciate valuable comments from Eugene Chiang and Scott Kenyon.  ANY thanks Phil Armitage for stimulating conversations. 

ANY acknowledges support from the {\it NASA } {\it ATP} and {\it OSS} grant NNX10AF35G and the  {\it NASA} {\it OPR} grant NNX11AM37G while at the CfA, and from the {\it NSF AST}  grant 1313021 and the {\it NASA OSS}  grant NNX13AI58G while at JILA.}

\bibliographystyle{apj}
\bibliography{refs}

\begin{thebibliography}{52}
\expandafter\ifx\csname natexlab\endcsname\relax\def\natexlab#1{#1}\fi

\bibitem[{{Alibert} {et~al.}(2005){Alibert}, {Mordasini}, {Benz}, \&
  {Winisdoerffer}}]{alibert05}
{Alibert}, Y., {Mordasini}, C., {Benz}, W., \& {Winisdoerffer}, C. 2005, \aap,
  434, 343

\bibitem[{{Andrews} {et~al.}(2010){Andrews}, {Wilner}, {Hughes}, {Qi}, \&
  {Dullemond}}]{andrews10}
{Andrews}, S.~M., {Wilner}, D.~J., {Hughes}, A.~M., {Qi}, C., \& {Dullemond},
  C.~P. 2010, \apj, 723, 1241

\bibitem[{{Arras} \& {Bildsten}(2006)}]{ab06}
{Arras}, P. \& {Bildsten}, L. 2006, \apj, 650, 394

\bibitem[{{Ayliffe} \& {Bate}(2012)}]{AylBat12}
{Ayliffe}, B.~A. \& {Bate}, M.~R. 2012, \mnras, 427, 2597

\bibitem[{{Bell} {et~al.}(2013){Bell}, {Naylor}, {Mayne}, {Jeffries}, \&
  {Littlefair}}]{bell13}
{Bell}, C.~P.~M., {Naylor}, T., {Mayne}, N.~J., {Jeffries}, R.~D., \&
  {Littlefair}, S.~P. 2013, \mnras, 434, 806

\bibitem[{{Bell} \& {Lin}(1994)}]{bell94}
{Bell}, K.~R. \& {Lin}, D.~N.~C. 1994, \apj, 427, 987

\bibitem[{{Bodenheimer} \& {Pollack}(1986)}]{boden86}
{Bodenheimer}, P. \& {Pollack}, J.~B. 1986, Icarus, 67, 391

\bibitem[{{Boss}(1997)}]{boss97}
{Boss}, A.~P. 1997, Science, 276, 1836

\bibitem[{{Bromley} \& {Kenyon}(2011)}]{bromley11}
{Bromley}, B.~C. \& {Kenyon}, S.~J. 2011, \apj, 731, 101

\bibitem[{{Burrows} {et~al.}(1997){Burrows}, {Marley}, {Hubbard}, {Lunine},
  {Guillot}, {Saumon}, {Freedman}, {Sudarsky}, \& {Sharp}}]{burrows97}
{Burrows}, A., {Marley}, M., {Hubbard}, W.~B., {Lunine}, J.~I., {Guillot}, T.,
  {Saumon}, D., {Freedman}, R., {Sudarsky}, D., \& {Sharp}, C. 1997, \apj, 491,
  856

\bibitem[{{Cameron}(1978)}]{cameron78}
{Cameron}, A.~G.~W. 1978, Moon and Planets, 18, 5

\bibitem[{{Chiang} \& {Youdin}(2010)}]{chiang10}
{Chiang}, E. \& {Youdin}, A.~N. 2010, Annual Review of Earth and Planetary
  Sciences, 38, 493

\bibitem[{{Close} {et~al.}(2014){Close}, {Follette}, {Males}, {Puglisi},
  {Xompero}, {Apai}, {Najita}, {Weinberger}, {Morzinski}, {Rodigas}, {Hinz},
  {Bailey}, \& {Briguglio}}]{close14}
{Close}, L.~M., {Follette}, K.~B., {Males}, J.~R., {Puglisi}, A., {Xompero},
  M., {Apai}, D., {Najita}, J., {Weinberger}, A.~J., {Morzinski}, K.,
  {Rodigas}, T.~J., {Hinz}, P., {Bailey}, V., \& {Briguglio}, R. 2014, \apjl,
  781, L30

\bibitem[{{D'Angelo} {et~al.}(2011){D'Angelo}, {Durisen}, \&
  {Lissauer}}]{dangelo11}
{D'Angelo}, G., {Durisen}, R.~H., \& {Lissauer}, J.~J. {Giant Planet
  Formation}, ed. S.~{Piper}, 319--346

\bibitem[{{Dones} \& {Tremaine}(1993)}]{dones93}
{Dones}, L. \& {Tremaine}, S. 1993, Icarus, 103, 67

\bibitem[{{Ferguson} {et~al.}(2005){Ferguson}, {Alexander}, {Allard}, {Barman},
  {Bodnarik}, {Hauschildt}, {Heffner-Wong}, \& {Tamanai}}]{FerAle05}
{Ferguson}, J.~W., {Alexander}, D.~R., {Allard}, F., {Barman}, T., {Bodnarik},
  J.~G., {Hauschildt}, P.~H., {Heffner-Wong}, A., \& {Tamanai}, A. 2005, \apj,
  623, 585

\bibitem[{{Fortney} {et~al.}(2007){Fortney}, {Marley}, \& {Barnes}}]{fortney07}
{Fortney}, J.~J., {Marley}, M.~S., \& {Barnes}, J.~W. 2007, \apj, 659, 1661

\bibitem[{{Hinz} {et~al.}(2012){Hinz}, {Arbo}, {Bailey}, {Connors}, {Durney},
  {Esposito}, {Hoffmann}, {Jones}, {Leisenring}, {Montoya}, {Nash}, {Nelson},
  {McMahon}, {Pinna}, {Puglisi}, {Skemer}, {Skrutskie}, \&
  {Vaitheeswaran}}]{hinz12}
{Hinz}, P., {Arbo}, P., {Bailey}, V., {Connors}, T., {Durney}, O., {Esposito},
  S., {Hoffmann}, W., {Jones}, T., {Leisenring}, J., {Montoya}, M., {Nash}, M.,
  {Nelson}, M., {McMahon}, T., {Pinna}, E., {Puglisi}, A., {Skemer}, A.,
  {Skrutskie}, M., \& {Vaitheeswaran}, V. 2012, in Society of Photo-Optical
  Instrumentation Engineers (SPIE) Conference Series, Vol. 8445, Society of
  Photo-Optical Instrumentation Engineers (SPIE) Conference Series

\bibitem[{{Hori} \& {Ikoma}(2010)}]{hori10}
{Hori}, Y. \& {Ikoma}, M. 2010, \apj, 714, 1343

\bibitem[{{Hori} \& {Ikoma}(2011)}]{HorIko11}
---. 2011, \mnras, 416, 1419

\bibitem[{{Hubickyj} {et~al.}(2005){Hubickyj}, {Bodenheimer}, \&
  {Lissauer}}]{HubBod05}
{Hubickyj}, O., {Bodenheimer}, P., \& {Lissauer}, J.~J. 2005, Icarus, 179, 415

\bibitem[{{Ida} {et~al.}(2013){Ida}, {Lin}, \& {Nagasawa}}]{ida13}
{Ida}, S., {Lin}, D.~N.~C., \& {Nagasawa}, M. 2013, \apj, 775, 42

\bibitem[{{Ikoma} {et~al.}(2000){Ikoma}, {Nakazawa}, \& {Emori}}]{ikoma00}
{Ikoma}, M., {Nakazawa}, K., \& {Emori}, H. 2000, \apj, 537, 1013

\bibitem[{{Kenyon} \& {Bromley}(2009)}]{kenyon09}
{Kenyon}, S.~J. \& {Bromley}, B.~C. 2009, \apjl, 690, L140

\bibitem[{{Kippenhahn} \& {Weigert}(1990)}]{kippenhahn90}
{Kippenhahn}, R. \& {Weigert}, A. 1990, {Stellar Structure and Evolution}

\bibitem[{{Kratter} {et~al.}(2010){Kratter}, {Murray-Clay}, \&
  {Youdin}}]{kratter10}
{Kratter}, K.~M., {Murray-Clay}, R.~A., \& {Youdin}, A.~N. 2010, \apj, 710,
  1375

\bibitem[{{Lagrange} {et~al.}(2010){Lagrange}, {Bonnefoy}, {Chauvin}, {Apai},
  {Ehrenreich}, {Boccaletti}, {Gratadour}, {Rouan}, {Mouillet}, {Lacour}, \&
  {Kasper}}]{lagrange10}
{Lagrange}, A.-M., {Bonnefoy}, M., {Chauvin}, G., {Apai}, D., {Ehrenreich}, D.,
  {Boccaletti}, A., {Gratadour}, D., {Rouan}, D., {Mouillet}, D., {Lacour}, S.,
  \& {Kasper}, M. 2010, Science, 329, 57

\bibitem[{{Lambrechts} \& {Johansen}(2012)}]{lambrechts12}
{Lambrechts}, M. \& {Johansen}, A. 2012, \aap, 544, A32

\bibitem[{{Lissauer} {et~al.}(2009){Lissauer}, {Hubickyj}, {D'Angelo}, \&
  {Bodenheimer}}]{LisHub09}
{Lissauer}, J.~J., {Hubickyj}, O., {D'Angelo}, G., \& {Bodenheimer}, P. 2009,
  Icarus, 199, 338

\bibitem[{{Macintosh} {et~al.}(2012){Macintosh}, {Anthony}, {Atwood},
  {Barriga}, {Bauman}, {Caputa}, {Chilcote}, {Dillon}, {Doyon}, {Dunn},
  {Gavel}, {Galvez}, {Goodsell}, {Graham}, {Hartung}, {Isaacs}, {Kerley},
  {Konopacky}, {Labrie}, {Larkin}, {Maire}, {Marois}, {Millar-Blanchaer},
  {Nunez}, {Oppenheimer}, {Palmer}, {Pazder}, {Perrin}, {Poyneer}, {Quirez},
  {Rantakyro}, {Reshtov}, {Saddlemyer}, {Sadakuni}, {Savransky},
  {Sivaramakrishnan}, {Smith}, {Soummer}, {Thomas}, {Wallace}, {Weiss}, \&
  {Wiktorowicz}}]{macintosh12}
{Macintosh}, B.~A., {Anthony}, A., {Atwood}, J., {Barriga}, N., {Bauman}, B.,
  {Caputa}, K., {Chilcote}, J., {Dillon}, D., {Doyon}, R., {Dunn}, J., {Gavel},
  D.~T., {Galvez}, R., {Goodsell}, S.~J., {Graham}, J.~R., {Hartung}, M.,
  {Isaacs}, J., {Kerley}, D., {Konopacky}, Q., {Labrie}, K., {Larkin}, J.~E.,
  {Maire}, J., {Marois}, C., {Millar-Blanchaer}, M., {Nunez}, A.,
  {Oppenheimer}, B.~R., {Palmer}, D.~W., {Pazder}, J., {Perrin}, M., {Poyneer},
  L.~A., {Quirez}, C., {Rantakyro}, F., {Reshtov}, V., {Saddlemyer}, L.,
  {Sadakuni}, N., {Savransky}, D., {Sivaramakrishnan}, A., {Smith}, M.,
  {Soummer}, R., {Thomas}, S., {Wallace}, J.~K., {Weiss}, J., \& {Wiktorowicz},
  S. 2012, in Society of Photo-Optical Instrumentation Engineers (SPIE)
  Conference Series, Vol. 8446, Society of Photo-Optical Instrumentation
  Engineers (SPIE) Conference Series

\bibitem[{{Marleau} \& {Cumming}(2013)}]{marleau13}
{Marleau}, G.-D. \& {Cumming}, A. 2013, arXiv:1302.1517

\bibitem[{{Marois} {et~al.}(2008){Marois}, {Macintosh}, {Barman}, {Zuckerman},
  {Song}, {Patience}, {Lafreni{\`e}re}, \& {Doyon}}]{marois08}
{Marois}, C., {Macintosh}, B., {Barman}, T., {Zuckerman}, B., {Song}, I.,
  {Patience}, J., {Lafreni{\`e}re}, D., \& {Doyon}, R. 2008, Science, 322, 1348

\bibitem[{{Matzner} \& {Levin}(2005)}]{matzner05}
{Matzner}, C.~D. \& {Levin}, Y. 2005, \apj, 628, 817

\bibitem[{{Menou} \& {Goodman}(2004)}]{menou04}
{Menou}, K. \& {Goodman}, J. 2004, \apj, 606, 520

\bibitem[{{Mizuno} {et~al.}(1978){Mizuno}, {Nakazawa}, \& {Hayashi}}]{mizuno78}
{Mizuno}, H., {Nakazawa}, K., \& {Hayashi}, C. 1978, Progress of Theoretical
  Physics, 60, 699

\bibitem[{{Ormel}(2013)}]{Orm13}
{Ormel}, C.~W. 2013, \mnras, 428, 3526

\bibitem[{{Ormel} \& {Klahr}(2010)}]{ormel10}
{Ormel}, C.~W. \& {Klahr}, H.~H. 2010, \aap, 520, A43

\bibitem[{{Papaloizou} \& {Nelson}(2005)}]{pn05}
{Papaloizou}, J.~C.~B. \& {Nelson}, R.~P. 2005, \aap, 433, 247

\bibitem[{{Papaloizou} \& {Terquem}(1999)}]{pap99}
{Papaloizou}, J.~C.~B. \& {Terquem}, C. 1999, \apj, 521, 823

\bibitem[{{Perri} \& {Cameron}(1974)}]{PerCam74}
{Perri}, F. \& {Cameron}, A.~G.~W. 1974, Icarus, 22, 416

\bibitem[{{Podolak}(2003)}]{podolak03}
{Podolak}, M. 2003, Icarus, 165, 428

\bibitem[{{Pollack} {et~al.}(1996){Pollack}, {Hubickyj}, {Bodenheimer},
  {Lissauer}, {Podolak}, \& {Greenzweig}}]{pollack96}
{Pollack}, J.~B., {Hubickyj}, O., {Bodenheimer}, P., {Lissauer}, J.~J.,
  {Podolak}, M., \& {Greenzweig}, Y. 1996, Icarus, 124, 62

\bibitem[{{Rafikov}(2005)}]{rafikov05}
{Rafikov}, R.~R. 2005, \apjl, 621, L69

\bibitem[{{Rafikov}(2006)}]{rafikov06}
---. 2006, \apj, 648, 666

\bibitem[{{Rafikov}(2011)}]{Raf11}
---. 2011, \apj, 727, 86

\bibitem[{{Saumon} {et~al.}(1995){Saumon}, {Chabrier}, \& {van
  Horn}}]{saumon95}
{Saumon}, D., {Chabrier}, G., \& {van Horn}, H.~M. 1995, \apjs, 99, 713

\bibitem[{{Semenov} {et~al.}(2003){Semenov}, {Henning}, {Helling}, {Ilgner}, \&
  {Sedlmayr}}]{semenov03}
{Semenov}, D., {Henning}, T., {Helling}, C., {Ilgner}, M., \& {Sedlmayr}, E.
  2003, \aap, 410, 611

\bibitem[{{Stevenson}(1982)}]{stevenson82}
{Stevenson}, D.~J. 1982, \planss, 30, 755

\bibitem[{{Williams} \& {Cieza}(2011)}]{williams11}
{Williams}, J.~P. \& {Cieza}, L.~A. 2011, \araa, 49, 67

\bibitem[{{Youdin} \& {Kenyon}(2013)}]{youdin13}
{Youdin}, A.~N. \& {Kenyon}, S.~J. {From Disks to Planets}, ed. T.~D. {Oswalt},
  L.~M. {French}, \& P.~{Kalas}, 1

\bibitem[{{Youdin} \& {Mitchell}(2010)}]{ym10}
{Youdin}, A.~N. \& {Mitchell}, J.~L. 2010, \apj, 721, 1113

\bibitem[{{Zhu} {et~al.}(2013){Zhu}, {Stone}, \& {Rafikov}}]{zhu13}
{Zhu}, Z., {Stone}, J.~M., \& {Rafikov}, R.~R. 2013, \apj, 768, 143

\end{thebibliography}

\appendix
\section{Derivation of the Global Energy Equation}\label{sec:globalderiv}

To derive  the global energy equation (\ref{eq:coolingglobal}) for an embedded protoplanet, we generalize the analogous calculations in stellar structure theory, e.g.\ in \S4.3 of \citet{kippenhahn90}.  For our problem, we add the effects of finite core radius, surface pressure and mass accretion. We start with the local energy equation (\ref{eq:structd}), whose more natural form in Lagrangian (mass) coordinates is $\p L/ \p m = \epsilon - T \p S /\p t$.  Integrating from the core to a higher shell with enclosed mass $M$ gives:
\begin{subeqnarray}
L - L\co &=& \int_{M\co}^M {\p L \over \p m} dm \\
&=& \int_{M\co}^M \left(\epsilon - T {\p S \over \p t} \right)dm \\
&=& \Gamma  - \int_{M\co}^M{\p u \over \p t} dm +  \int_{M\co}^M {P \over \rho^2} {\p \rho \over \p t} dm\slabel{eq:DLc}\, ,
\end{subeqnarray} 
with $\Gamma = \int \epsilon dm$ the integral of the direct heating rate, and applying the first law of thermodynamics in the final step.

The global energy equation is derived by eliminating the partial time derivatives in \Eq{eq:DLc}, which are performed at a fixed mass,
in favor of total time derivatives, denoted with overdots.  
For instance, the surface radius $R$ of the shell with enclosed mass $M$ evolves as  
\begin{equation}\label{eq:Rdot}
 \dot{R} = {\p R \over \p t} + {\dot{M} \over 4 \pi R^2 \rho_M},
\end{equation} 
where $\p R/\p t$ gives the Lagrangian contraction of the ``original" shell, and mass accretion through the upper boundary at rate $\dot{M}$ also changes the shell location.  
Similarly, the volume $V = (4 \pi/3)R^3$ and pressure at the outer shell evolve as
\begin{subeqnarray}\label{eq:dot}
\dot{V}_M &=&  {\p V_{\rm M} \over \p t} + {\dot{M} \over \rho_{\rm M}}  \\
 \dot{P}_M &=& {\p P_{\rm M} \over \p t} + {\p P_M \over \p m}\dot{M} =  {\p P_{\rm M} \over \p t} - {G M  \over 4 \pi R^4} \dot{M}\, .
\end{subeqnarray} 
This derivation holds the core mass and radius fixed, $\dot{M}\co = \dot{R}\co = 0$.  Therefore, the core pressure satisfies
\begin{equation}\label{eq:Pcdot}
 \dot{P}\co = \p P\co / \p t \, .
\end{equation}
The internal energy integral follows simply from  Leibniz's rule as
\begin{equation}\label{eq:udot}
\int_{M\co}^{M(t)}{\p u \over \p t} dm = \dot{U}  -  \dot{M}u_M\, .
\end{equation} 
To make further progress, we use the virial theorem:
\begin{equation}
\label{eq:virial}
E_G=-3 \int_{M\co}^M \frac{P}{\rho} dm + 4 \pi (R^3 P_M-R\co^3 P\co),
\end{equation}
which follows from \Eqsss{eq:structb}{eq:structa}{eq:Eg} by integrating hydrostatic balance in Lagrangian coordinates.  As an aside, the integral in equation (\ref{eq:virial}) can be evaluated for a polytropic EOS to give simple expressions for the total energy:
\begin{subeqnarray}
E&=&(1-\zeta)U+4 \pi (R^3 P_M-R\co^3 P\co) \slabel{eq:vira} \\
&=&\frac{\zeta-1}{\zeta}E_G+\frac{4 \pi}{\zeta} (R^3 P_M-R\co^3 P\co) \slabel{eq:virb} \, ,
\end{subeqnarray}
where $\zeta \equiv 3(\gamma - 1)$.  We will not make this assumption and will keep the EOS general.

To express the work integral, i.e. the final term in \Eq{eq:DLc}, in terms of changes to gravitational energy, we first take the
 time derivative of \Eq{eq:virial}:
\begin{eqnarray}\label{eq:EGdot}
\dot{E}_G = 3  \int_{M\co}^M {P \over \rho^2} {\p \rho \over \p t} dm -3 \int_{M\co}^M {\p P\over \p t}{dm \over \rho} 
 -  3{P_M \over \rho_M} \dot{M}+ 3 \dot{P}_M V_M -3 \dot{P}\co V\co  + 3  P_M {\dot{ V}_M} \, . 
\end{eqnarray} 
The first integral in \Eq{eq:EGdot} is the one we want, but the second one must be eliminated.  The time derivative of \Eq{eq:Eg} (times four) gives
\begin{subeqnarray}
 4 \dot{E}_G &=&  -4 {G M \dot{M} \over R} + 4 \int_{M\co}^M {G m \over r^2}{\p r \over \p t} dm\\ 
&=&   -4 {G M \dot{M} \over R} + 4 \pi \int_{M\co}^M r^3{\p \over \p m}{\p P \over \p t} dm \slabel{eq:4EGb} \\
&=&  -4 {G M \dot{M} \over R} -3  \int_{M\co}^M {\p P\over \p t}{dm \over \rho}  + 3 V_M {\p P_M \over \p t} -3 V\co {\p P\co \over \p t} \slabel{eq:4EGc} \, ,
\end{subeqnarray} 
where \Eqs{eq:4EGb}{eq:4EGc} use hydrostatic balance  and integration by parts.


Subtracting \Eqs{eq:udot}{eq:4EGc} and rearranging terms with the help of \Eqsss{eq:Rdot}{eq:dot}{eq:Pcdot} gives
\begin{eqnarray}\label{eq:PdVint}
\int_{M\co}^M {P \over \rho^2} {\p \rho \over \p t} dm  &=&  - \dot{E}_G - {G M \dot{M} \over R} - P_M {\p V_M \over \p t} \,  .
\end{eqnarray} 
Combining \Eqsss{eq:DLc}{eq:udot}{eq:PdVint}, we reproduce \Eq{eq:coolingglobal} with the accreted specific energy $e_M \equiv u_M - GM/R$.  

\section{Analytic Cooling Model Details}\label{sec:analytic}

\subsection{Isothermal Atmosphere}
\label{iso}

We consider the structure of a non-self-gravitating, isothermal atmosphere that extends outward from the radiative-convective boundary (RCB) and matches onto the disk density, $\rho_{\rm d}$, at a distance $r_{\rm fit} = n_{\rm fit} \RB$, where $R_{\rm B}$ is the Bondi radius defined in equation (\ref{eq:RB}). From equation (\ref{eq:structa}) the resulting density profile is
\begin{equation} \label{eq:rhoiso}
\rho = \rho_{\rm d} \exp \left({R_{\rm B} \over r} - {1 \over n_{\rm{fit}}} \right) \approx   \rho_{\rm{d}} \exp \left(R_{\rm B} \over r  \right),
\end{equation} 
where the approximate inequality is valid deep inside the atmosphere ($r \ll \RB$) for any $n_{\rm fit} \gtrsim 1$.  However, the choice of boundary condition does have an order unity effect on the density near the Bondi radius. 

The mass of the atmosphere is determined by integrating equation (\ref{eq:structb}) from the RCB to the Bondi radius using the density profile (\ref{eq:rhoiso}) and can be approximated as
\begin{equation} \label{eq:MatmISO}
M_{\rm iso} \approx 4 \pi \rho\di {R\cb^4 \over R_{\rm B}} e^{R_B/R\cb} = 4\pi \rho\cb \frac{R\cb^4}{R_{\rm B}} \, ,
\end{equation}
with $\rho\cb$ the density at the RCB.
This result is the leading order term in a series expansion. By comparing the expression above and \Eq{eq:Matman} under the assumption that $R\cb \ll R_{\rm B}$, we see that the mass of the outer radiative region (which is nearly isothermal) is negligible when compared with the atmosphere mass in the convective layer, as stated in Section \ref{sec:coolingan}.

\subsection{Temperature and Pressure Corrections at the Radiative-Convective Boundary}
\label{RCBcorr}

We estimate the temperature and pressure corrections at the RCB due to the fact that the radiative region is not purely isothermal. From equation (\ref{eq:delrad}), we express the radiative lapse rate
\begin{equation}\label{eq:delradan}
\delrad = {3 \kappa P \over 64 \pi  G M \sigma T^4} L = \nabla\di {P/P_{\rm d} \over (T/T_{\rm d})^{4-\beta}},
\end{equation}

\noindent where the second equality follows from the opacity law (\ref{eq:opacitylaw}) and $\nabla_{\rm d}$ is the radiative temperature gradient at the disk:

\begin{equation}
\label{eq:delo}
\nabla \di \equiv \frac{3 \kappa(T{\di}) P{\di}}{64 \pi G M \sigma T_{\rm d}^4} L.
\end{equation}
Here $M$ is the total planet mass. Since our analytic model neglects self-gravity, $M=M\co$ and therefore $\nabla\di$ is constant. From equation (\ref{eq:delradan}) and $\delrad=d \ln T / d\ln P$, the temperature profile in the radiative region integrates to
\begin{equation}\label{eq:radTP}
\left(T \over T_{\rm d}\right)^{4-\beta} - 1 = {\nabla\di \over \nabla_\infty} \left( {P \over P_{\rm d}} - 1 \right) \, ,
\end{equation} 
where $\nabla_\infty = 1/(4-\beta)$ is the radiative temperature gradient for $T ,P \rightarrow \infty$.
Applying \Eqs{eq:delradan}{eq:radTP} at the RCB (where $\delrad = \delad$) under the assumption that $P\cb \gg P_{\rm d}$ results in  $T\cb=\chi T\di$ as in \Eq{eq:Tcb}, with $\chi$ defined in \Eq{eq:chi}.

The pressure at the RCB follows from \Eqs{eq:radTP}{eq:Tcb} as
\begin{equation}
\label{eq:Pcbapprox}
{P\cb\over P_{\rm d}} \simeq {\delad/\nabla\di \over 1 - \delad/\nabla_\infty}.
\end{equation} 
 We can eliminate $\nabla\di$ from equation (\ref{eq:Pcbapprox}) to obtain a relation between temperature and pressure in the radiative zone as a function of the RCB pressure $P\cb$. From \Eq{eq:radTP}, it follows that
 \begin{equation}\label{eq:TP}
{T \over T_{\rm d}} = \left[1 + {1 \over {\nabla_\infty \over \delad} - 1} \left({P \over P\cb} -  {P_{\rm d} \over P\cb}\right) \right]^{1 \over 4-\beta}\, .
\end{equation} 
 We can then determine the RCB radius $R\cb$ from \Eq{eq:structa} as 
\begin{equation}\label{eq:RCBint}
{R_B \over R\cb} = \int_{P\di}^{P\cb} {T \over T_{\rm d}} {dP \over P}\, .
\end{equation}
Evaluating the integral leads to 
\begin{equation}\label{eq:Rcb}
{R_B \over r\cb} = \ln \left(P\cb \over P\di \right) - \ln \theta \, ,
\end{equation} 
with an extra correction term $\theta < 1$, when compared to an isothermal atmosphere (see Equation \ref{eq:rhoiso}). From this we arrive at the relation between $P\cb$ and $P\di$ given by Equation (\ref{eq:PcbRcb}). As opposed from the temperature correction factor $\chi$, an analytic expression for  $\theta$ cannot be obtained. Estimates for $\chi$ and $\theta$ for different values of the exponent $\beta$ in the opacity law (\ref{eq:opacitylaw}) are presented in Table 1.



\subsection{The Opacity Effect}
\label{opacityan}
A  lower opacity  decreases the critical core mass.  Reducing the opacity by a factor of one hundred results in a critical core mass one order of magnitude lower than in the standard ISM case, for our analytic model. The reduction is not  as strong as the nominal scaling would imply, $0.01^{3/5} \approx 0.06$, because $\xi$ increases.

Even with significantly lower opacities, radiative diffusion remains a good approximation at the RCB. For $\beta = 2$, we estimate the optical depth as
\begin{equation}
\tau\cb \sim {\kappa\cb P\cb \over g} \sim 7 \times 10^4 {F_T^4 F_\kappa \over \left(\mc \over 10 \right) \left(\au \over 10\right)^{12 \over 7}}, 
\end{equation} 
where $P\cb \sim P_M$ for a self-gravitating atmosphere and $g \sim G M\co/R_B^2$, with both approximations good to within the order unity factor $\xi$.  We see that $\tau\cb \gg 1$ even for $F_\kappa \lesssim 0.01$ out to very wide separations, hence the atmosphere remains optically thick at the RCB.


\subsection{Surface Terms}
\label{surfterms}
In this section we check the relevance of the neglected surface terms in \Eq{eq:coolingglobal}.  We first show that accretion energy is only a small correction at the RCB, which is where we apply our cooling model. A rough comparison (ignoring terms of order $\xi$) of  accretion luminosity vs. $\dot{E}$ gives
\begin{equation}
{G M \dot{M} \over R \dot{E}} = {G M  \over R}{dM \over dE} \sim {G M\co^2 \over R_B E}{P\cb \over P_M} \sim \sqrt{R\co \over R_B} \ll 1,
\end{equation} 
where we assume $P\cb \sim P_M$  for a massive atmosphere.  Accretion energy at the protoplanetary surface is thus very weak for marginally self-gravitating atmospheres, and even weaker for lower mass atmospheres.  A similar scaling analysis shows that the work term $P_M \p V_M/\p t$ is similarly weak.  Nevertheless, our numerical calculations include these surface terms in a more realistic and complete model of self-gravitating atmospheres.

\begin{deluxetable}{cccccc}  
\gdef \numcols {6}
\tablecolumns{\numcols}
\tablecaption{Parameters Describing Structure of Radiative Zone.}
\tablehead{   \multicolumn{\numcols}{c} {$\gamma = 7/5$ ($\delad = 2/7$)} }  
\startdata
 $\beta$   		 &1/2  	& 3/4 &1   		& 3/2  		& 2   \\
 $\nabla_\infty$ & 2/7 \tablenotemark{a}  	&  4/13	& 1/3 	& 2/5 	 	& 1/2 \\
 $\chi$ 		 & \nodata &  2.25245 &1.91293 	& 1.65054 	& 1.52753 \\
 $\theta $  		 &\nodata   & 0.145032	&0.285824   &0.456333   & 0.556069   \\
\enddata
\tablenotetext{a}{Since $\delad = \nabla_\infty$ there is no convective transition at depth for this case.}
\end{deluxetable}

%

\end{document}